\documentclass[
 reprint,
 superscriptaddress,
 amsmath,amssymb,
aps,
pub,
]{revtex4-2}

\usepackage{physics}
\usepackage[utf8]{inputenc}
 
\usepackage{hyperref}
\usepackage{graphicx}
\usepackage{rotating}
\usepackage{amsmath,amssymb,amsfonts,textcomp}
\usepackage{color}
\usepackage{siunitx}

\usepackage{xspace}

\begin{document}

\title{Computing Exchange Coupling constants in Transition metal complexes with Tensor Product Selected Configuration Interaction}

\author{Arnab Bachhar}
\author{Nicholas J. Mayhall}
\email{nmayhall@iu.edu}
\affiliation{Department of Chemistry, Indiana University,
Bloomington, IN 47401, USA}

\begin{abstract}
Transition metal complexes present significant challenges for electronic structure theory due to strong electron correlation arising from partially filled $d$-orbitals. We compare our recently developed Tensor Product Selected Configuration Interaction (TPSCI) with Density Matrix Renormalization Group (DMRG) for computing exchange coupling constants in six transition metal systems, including dinuclear Cr, Fe, and Mn complexes and a tetranuclear Ni-cubane. TPSCI uses a locally correlated tensor product state basis to capture electronic structure efficiently while maintaining interpretability.
From calculations on active spaces ranging from (22e,29o) to (42e,49o), we find that TPSCI consistently yields higher variational energies than DMRG due to truncation of local cluster states, but provides magnetic exchange coupling constants (J) generally within 10-30 cm$^{-1}$ of DMRG results. Key advantages include natural multistate capability enabling direct J extrapolation with smaller statistical errors, and computational efficiency for challenging systems. However, cluster state truncation represents a fundamental limitation requiring careful convergence testing, particularly for large local cluster dimensions. We identify specific failure cases where current truncation schemes break down, highlighting the need for improved cluster state selection methods and distributed memory implementations to realize TPSCI's full potential for strongly correlated systems.
\end{abstract}

\maketitle
\section{Introduction}\label{sec:introduction}
Transition metal complexes (TMCs) are critically important for a wide range of chemical processes, such as catalysis,~\cite{oec,nitrogenase} electron transfer,~\cite{energy_bimetallic} molecular magnetism,~\cite{molecular_magnetism} energy storage, biological functions e.g. oxygen transport in hemoglobin, photooxidation of water,~\cite{Gil-Sepulcre2022}, nitrogen fixation,~\cite{TANABE2022214783} etc.
Modeling TMCs in computational chemistry is challenging due to their complex electronic structures and diverse properties arising from partially filled d-orbitals, which often lead to a dense low-energy spectrum and strong electron correlation, 
making the Hartree-Fock (HF) approximation~\cite{szabo__attila_modern_nodate,helgaker_molecular_2000-1,hartree_wave_1928} a poor reference for post-HF single-reference methods.~\cite{shavitt_many-body_2009, korth_density_2017}  
As a result, accurate qualitative predictions and quantitative accuracy of spin states and magnetic properties often require multiconfigurational methods, which generally scale combinatorially with system size, limiting active space sizes to less than around 20 orbitals in practice.~\cite{cramer2002essentials,li_block-correlated_2004,gasscf,cmf_first,parker_communication_2013,parkerCommunicationActiveSpace2014,vlasscf_hermes,vlasscf_dmet,parkerModelHamiltonianAnalysi2014}

The density matrix renormalization group (DMRG) algorithm has emerged as a state-of-the-art method for the computational study of strongly correlated systems, offering significant advantages over traditional electronic structure approaches, especially for quasi-1D systems.~\cite{pantazisMeetingChallengeMagnetic2019,szalayTensorProductMethod2015,
baiardiDensityMatrixRenormalization2020,
schollwockDensitymatrixRenormalizationGroup2011,woutersDensityMatrixRenormalization2014,
eriksenGroundStateElectronic2020,
olivares-amayaAbinitioDensityMatrix2015a,
sharmaLowenergySpectrumIronsulfur2014, 
sharmaQuasidegeneratePerturbationTheory2016a}
In DMRG, the matrix product state (MPS) ansatz provides the ability to capture
complex entanglement between orbitals without bias for any particular reference state. DMRG's ability to treat static correlation accurately, extendable to the FCI limit within the chosen active space, makes it effective for TMCs.
Despite the fact that DMRG is formally ideal for 1D systems, it has shown great utility in applications to even compact nonlinear systems such as multi-centered TMCs and clusters with large active space and larger bond dimension ($\chi$) of the MPS.~\cite{zhai2021low}

More recently, we have proposed Tensor Product Selected Configuration Interaction (TPSCI) as a promising approach for studying strongly correlated systems and excited states.~\cite{abraham_selected_2020,braunscheidel_generalization_2023,faraday_discussions} 
TPSCI leverages a locally correlated tensor product state (TPS) basis to capture electronic structure in complex molecular systems efficiently.
TPSCI addresses the challenge of computing accurate yet interpretable wavefunctions by concentrating the weight of a state onto a smaller number of physically meaningful degrees of freedom. 
We have demonstrated TPSCI's ability to provide accurate solutions for challenging TMCs and other strongly correlated systems while maintaining computational efficiency.~\cite{abraham_selected_2020,braunscheidel_generalization_2023,faraday_discussions}  
The method builds on concepts from cluster mean-field theory (cMF)~\cite{cmf_first,cmf_third,cMF_linear_combinations,bachhar_rocmf} 
Active-Space Decomposition (ASD)~\cite{parker_communication_2013,parkerCommunicationActiveSpace2014} and selected CI (SCI) algorithms,~\cite{huron1973iterative,evangelisti1983convergence,loos2018mountaineering}, combining them to create a powerful tool for analyzing complex electronic systems. 
While TPSCI has been shown to provide advantages compared to Slater determinant-based SCI methods for large active spaces of closed-shell systems, 
 no benchmark calculations have been reported yet for open-shell systems, such as TMC's. 

In this paper, we present a comprehensive study comparing the performance of TPSCI against DMRG for a series of TMC's.
This study systematically evaluates the ability of TPSCI to handle large active spaces and predict relative energetics of spin states, and benchmarking these results against the highly accurate yet computationally demanding DMRG.
We will mainly focus on relative spin-state energies, exchange coupling constant between magnetic centers, and the standard errors for extrapolation of variational energies.

By highlighting the strengths and limitations of TPSCI relative to DMRG, this work aims to establish its utility and reliability as an alternative method for tackling the challenging electronic structure of TMCs.
A key focus of this work is to identify the weaknesses and shortcomings of TPSCI to help direct future development for more effective simulations of complex systems. 

\section{Theory}\label{theory}

\subsection{Tensor product space}
We start by choosing an active space sufficient to include all the necessary physics. The second quantized electronic Hamiltonian can be expressed in terms of active set orbitals, $(p,q,r,s)$,
\begin{align}
    \hat{H} = h_{pq}\hat{p}^\dagger\hat{q} + \tfrac{1}{2}\left<pq|rs\right>\hat{p}^\dagger\hat{q}^\dagger\hat{s}\hat{r},
\end{align}
and choosing an optimal such set is an important aspect of all active space methods (including both TPSCI and DMRG).
In fact, considering comparison to experiment, the active spaces used in this paper (despite being large) is one of the most limiting assumptions in this article.

To achieve a compact and interpretable wavefunction representation, the active orbitals can be divided into separate clusters or groups.
These clusters, denoted by capital letters such as \textbf{I}, are organized to maximize the interactions within each cluster while minimizing those between different clusters.
For bimetallic compounds, we define each cluster to consist of all orbitals localized around a specific metal center.
This approach ensures that intra-cluster correlations capture the local dynamical interactions entirely, while weaker inter-cluster effects, such as spin coupling, are treated separately.

Each orbital cluster defines a local Hilbert space, within which we construct a local basis of cluster states such that each cluster state is a linear combination of all possible Slater determinants within that particular cluster. 
The entire global Hilbert space can be accessed by forming all possible tensor products of local cluster states ($|\alpha\rangle, |\beta\rangle, \cdots$). A global tensor product state (TPS) can be expressed using these local many-body cluster states  as:
\begin{align}
    |\Phi\rangle=|\alpha\rangle_\textbf{I} |\beta\rangle_\textbf{J} \cdots |\delta\rangle_\textbf{N},
\end{align}
where $|\alpha\rangle$ spans the entire Fock space of cluster $\textbf{I}$.

Because all local correlation can be included inside the cluster states themselves, a single TPS basis can incorporate a significant amount of electron correlation. Consequently, the wavefunction written in terms of the TPS basis can often require far fewer basis vectors than in the Slater determinant basis analog. The exact wavefunction is expressed in terms of composite states formed by the tensor product of cluster states,
\begin{align}\label{eq:tps_presentation}
|{\psi}\rangle =& \sum_{\textbf{n}}\sum_{i\in n_\textbf{I}}\sum_{j\in n_\textbf{J}}\dots\sum_{n\in n_\textbf{N}}
C^{n_\textbf{I}, n_\textbf{J}, \dots, n_\textbf{N}}_{i,j,\dots,n}|{\alpha_i^{n_\textbf{I}}}\rangle |{\beta_j^{n_\textbf{J}}}\rangle \cdots|{\delta_n^{n_\textbf{N}}}\rangle,
\end{align}
where $i, j, \cdots, n$ represent Slater determinants localized on cluster $\textbf{I}, \textbf{J}, \cdots, \textbf{N}$, respectively. $C$ is the coefficient tensor, and this is grouped into different blocks, each of which represents a specific local quantum number string, \(\textbf{n} = \{n_1, n_2, \ldots, n_N\}\).
Notably, this encompasses cluster states with different numbers of particles. Although the formalism presented is general, optimal performance is typically achieved when the interactions within a cluster are stronger than those between clusters.

The total Hamiltonian, $\hat{H}$ can be divided into contributions based on the number of distinct clusters involved: one-, two-, three-, and four-cluster terms. These contributions are defined as follows:
\begin{align}\label{eq:ham}
    \hat{H} = & \sum_\textbf{I} \hat{H_\textbf{I}} + \sum_{\textbf{I}<\textbf{J}} \hat{H}_{\textbf{IJ}} \nonumber\\
    &+ \sum_{\textbf{I}<\textbf{J}<\textbf{K}} \hat{H}_{\textbf{IJK}} + \sum_{\textbf{I}<\textbf{J}<\textbf{K}<\textbf{L}}\hat{H}_{\textbf{IJKL}}.
\end{align}

In Eq. (\ref{eq:ham}), $\hat{H_\textbf{I}}$ includes terms where all creation and annihilation operators are within cluster $\textbf{I}$, $\hat{H}_{\textbf{IJ}}$ involves operators from both clusters $\textbf{I}$ and $\textbf{J}$, and so on.
As the ab initio Hamiltonian consists solely of two-body interactions, the maximum number of clusters involved in the interactions is limited to four only.

\subsection{cluster Mean-Field Theory}
As defined in Eq. \ref{eq:tps_presentation}, the local symmetries, including spin projection and particle number, remain conserved for each cluster state.
This cluster formalism takes care of the local correlations, i.e., intra-cluster correlations only, but neglects the inter-cluster correlations.
A significant improvement can be achieved by defining the local cluster states in the averaged field of the remaining clusters, using the Cluster Mean-Field (cMF) theory.~\cite{cmf_first, bachhar_rocmf}
In cMF, the local cluster states are defined as the eigenstates of an effective cluster Hamiltonian:
\begin{align}\label{eq:cmf_hamiltonian}
\hat{H}^{cMF}_\textbf{I} =& 
\sum_{pq\in \textbf{I}}h_{pq} \hat{p}^\dagger \hat{q}+\sum_{pqrs\in \textbf{I}}\left<pq|rs\right> \hat{p}^\dagger\hat{q}^\dagger\hat{s}\hat{r} \\ \nonumber
&+ \sum_{\textbf{J}\neq \textbf{I}}\sum_{pq\in \textbf{I}}\sum_{rs\in \textbf{J}}\left<pr||qs\right>\gamma^{\textbf{J}}_{rs}\hat{p}^\dagger \hat{q},
\end{align}
where $\gamma^{\textbf{J}}_{rs}=\left<\beta_0^{n_\textbf{J}}|\hat{r}^\dagger\hat{s}|\beta_0^{n_\textbf{J}}\right>$, represents the one-particle reduced density matrix (1RDM) of the ground state on cluster $\mathbf{J}$.
The third term in Eq.(\ref{eq:cmf_hamiltonian}) accounts for the mean-field interactions of cluster $\mathbf{I}$ with the rest of the system.
Moreover, since the mean-field term is computed using the ground state densities of the clusters, this process can be seen as a variational minimization to obtain the most optimal single tensor product state. This is analogous to how the Hartree-Fock theory achieves the best approximation with a single Slater determinant. Similar to CASSCF, the cMF energy is minimized with respect to both the coefficients of the cluster states and also with respect to global inter-cluster orbital rotation parameters.

Unrestricted cMF (UcMF)~\cite{cmf_first} treats clusters containing unpaired electrons by allowing the $\alpha$ and $\beta$ orbitals to spin polarize, leading to significant energetic stabilization.
However, this naturally results in a spin-contaminated cMF wavefunction, making it challenging to use as a reference state for spin-pure post-cMF methods.
Recently, we introduced a spin-preserving version of cMF (restricted open-shell cMF (RO-cMF)~\cite{bachhar_rocmf}), which extends the cMF framework to systems with open-shell clusters while constraining spin polarization.
Although RO-cMF necessarily yields higher energies than the UcMF approach, it provides a well-defined, spin-pure reference state for post-cMF methods that account for missing inter-cluster correlations.

Although cMF provides a qualitatively accurate description of the ground state, it lacks quantitative accuracy due to missing inter-cluster interactions.
We define a cluster basis that involves constructing a set of many-electron basis states for each cluster, typically including important configurations that capture the essential physics of the system.~\cite{abraham_selected_2020,braunscheidel_generalization_2023,faraday_discussions,abraham_cluster_2021,abraham_coupled_2022}
The cMF calculation yields not only the variationally optimal tensor product state but also a set of cluster-local effective Hamiltonians, each incorporating mean-field interactions from the surrounding clusters. The eigenvectors of these effective Hamiltonians are used as the initial basis for the cluster states.
We can vary a parameter $\delta_e$ for change in particle number in each of the clusters, and in this way, we can control the computational cost by not including irrelevant Fock sectors that contribute negligibly to the preferable final TPSCI wavefunction.
Once the allowed particle number subspaces (Fock sectors) are defined, the eigenvectors of the local cMF effective Hamiltonians are computed within each cluster.
A total of $M$ (a user-defined parameter) eigenvectors are obtained for each Fock sector and stored as basis vectors.
Since these eigenvectors diagonalize a local Hamiltonian, all local correlations are effectively incorporated into the basis representation.

Truncation of local Fock sectors to $M$ states can generate spin contamination if not done carefully.
To avoid spin contamination, we generate the high- and low-$ m_S $ components by directly applying the spin raising $ S^+ $ and lowering $S^- $ operators to the $m_S = 0 $ eigenstates for clusters with an even number of electrons and to the $m_S = \frac{1}{2} $ eigenstates for clusters with an odd number of electrons.~\cite{braunscheidel_generalization_2023}
This approach ensures that all $ m_S $ components are included for each computed cluster state, preserving $\hat{S}_z $ symmetry when truncating the number of states to $M$.
The value of $M$ (maximum number of cluster states in each cluster Fock space) controls the number of states to include in the cluster basis for a particular Fock sector and thus keeps a check in computational cost by truncating the Hilbert space of a particular active space.

\subsection{Tensor product selected CI}
Once the cMF wavefunction has been converged, computing the higher energy eigenvectors of the cluster mean-field Hamiltonians, $\hat{H}^{cMF}_\textbf{I}$, provides a natural TPS orthonormal basis for the entire Hilbert space.
If no truncation is done at this step (meaning all eigenvectors of $\hat{H}^{cMF}_\textbf{I}$ are computed), then the dimension of the Hilbert space is unchanged and equally difficult to solve.
A key advantage of employing a TPS basis with diagonal local Hamiltonians is that local correlations are inherently incorporated into the basis states themselves.
As a result, low-energy solutions tend to be more concentrated upon a smaller number of basis states.
This can improve the efficiency of computational methods that leverage sparsity, such as selected CI, making them potentially more effective in the correlated TPS basis compared to a Slater determinant basis.
We have already shown in Refs. ~\cite{abraham_selected_2020, braunscheidel_generalization_2023} that this often holds and sometimes provides us computational benefit.

In tensor product selected CI (TPSCI), the CIPSI~\cite{huron1973iterative} algorithm is generally employed to exploit the sparsity of wavefunction in a TPS basis.
It follows certain steps:
\begin{enumerate}
    \item We build and diagonalize the Hamiltonian in the current variational ($\mathcal{P}$) space (a set of TPS states that should have large contributions towards the exact solution).
    \item Apply the Hamiltonian to the current variational eigenvectors to couple them with the external ($\mathcal{Q}$) space.
    \item We search throughout the $\mathcal{Q}$ space using a perturbation theory (defined by using Löwdin’s partitioning theory) and form first-order interaction space.
    \item Add external configurations with larger first-order coefficients than a threshold to the $\mathcal{P}$-space.
    \item Repeat steps, 1-4 if the dimension of the $\mathcal{P}$-space keeps increasing. If not, exit as TPSCI energy is converged.
\end{enumerate}
\begin{figure}
    \centering
    \includegraphics[width=0.8\linewidth]{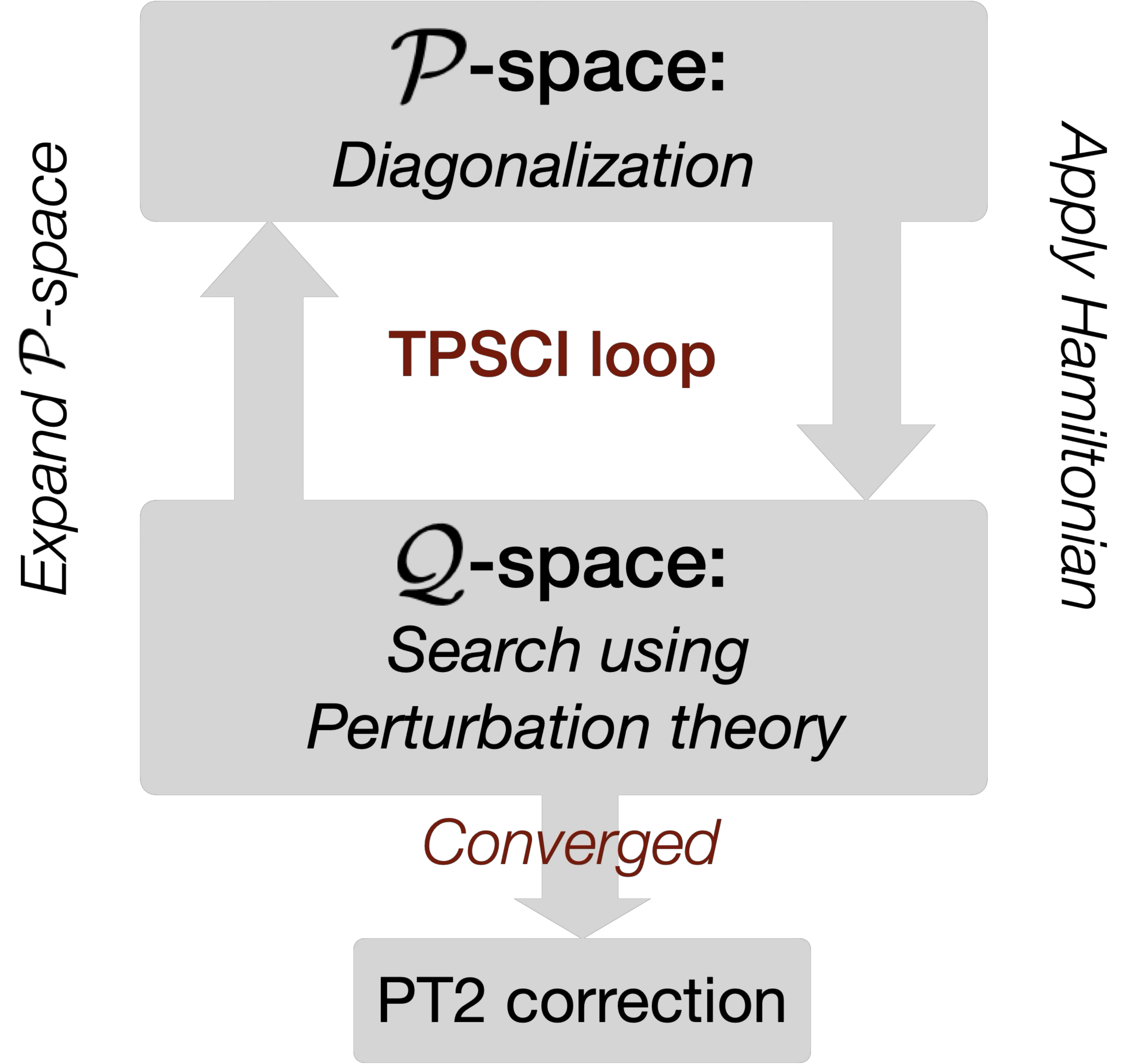}
    \caption{Schematic illustration of a TPSCI algorithm. TPSCI loop is iterated until the dimension of the $\mathcal{P}$-space (variational space) stops growing. After the TPSCI energy is converged, a batched PT2 energy correction is done over the TPSCI wave function to get the missing dynamical correlation in the active space.}
    \label{fig:cipsi}
\end{figure}

One of the important steps of TPSCI is to construct matrix elements of the Hamiltonian.
For example, one of the many two-cluster terms of the Hamiltonian acting between two arbitrary TPS configurations, 
$|\Psi\rangle$ and $|\Psi^{\prime}\rangle$, can be expressed as:
\begin{align}
\hat{H}_{\textbf{I}\textbf{J} }=&\sum_{p r}^{\textbf{I}} \sum_{q s}^{\textbf{J}}\langle p q \| r s\rangle\left\{\hat{p}^{\dagger}  \hat{r}\right\}_\textbf{I}\{\hat{q}^{\dagger}\hat{s}\}_\textbf{J},\\
\left\langle\Psi^{\prime}\left|\hat{H}_{\textbf{I} \textbf{J}}\right| \Psi\right\rangle =&-(-1)^{\zeta} \sum_{p r \in \textbf{I}} \sum_{q s \in \textbf{J}}\langle p q \| r s\rangle \Gamma_{p r}^{\beta^{\prime} \beta} \Gamma_{qs}^{\gamma^{\prime} \gamma} \nonumber\\&\times  \prod_{\textbf{K} \neq \textbf{I}, \textbf{J}}\delta_{\omega_\textbf{K},\omega^\prime_\textbf{K}}
,\\
\Gamma_{p r}^{\beta^{\prime} \beta}=&\left\langle\beta^{\prime}\left|\hat{p}^{\dagger} \hat{r}\right| \beta\right\rangle_{\textbf{I}},\\
  \Gamma_{q s}^{\gamma^{\prime} \gamma}=&\left\langle\gamma^{\prime}|\hat{q}^{\dagger} \hat{s}| \gamma\right\rangle_{\textbf{J}},
  \end{align}
where $\alpha, \beta, \cdots $ are correlated many-body states on clusters, 
and $\zeta=\sum_{\textbf{K}=\textbf{I}}^{\textbf{J-1}} N_{\textbf{K}}$ is the electron count in each state on clusters located between the two active clusters, 
and where the gamma tensors, $\Gamma_{p r}^{\beta^{\prime}\beta}$ and $\Gamma_{q s}^{\gamma^{\prime} \gamma}$ are specific to cluster \textbf{I} and \textbf{J}, respectively.
 The term $\delta_{\omega_\textbf{K},\omega^\prime_\textbf{K}}$ represents orthonormality condition between states $\omega$ and $\omega^\prime$ for cluster \textbf{K}, where $\omega $ belongs to specific state in wavefunction $|\Psi\rangle$.
The $\Gamma$ tensors are the precomputed local operator matrices in the cluster basis and are contracted with the integrals to compute the matrix elements. 

The construction of matrix elements becomes a key bottleneck in the TPSCI algorithm, requiring a series of tensor constructions for each matrix element. 
This amounts to a significant space to time tradeoff, one that is helpful when traditional Selected CI methods run out of memory.
Previously, we have compared TPSCI\cite{abraham_selected_2020} with heat bath CI (HCI)\cite{holmesHeatBathCI2016} and were able to obtain significantly lower TPSCI variational energy than with HCI.

Once the TPSCI variational space has been sufficiently converged for a particular threshold, an additional refinement of cluster basis can sometimes be beneficial, particularly for ground-state calculations. Specifically, applying a higher-order singular value decomposition (HOSVD) to the wavefunction tensor allows for a rotation of cluster states, optimizing them into a form that diagonalizes the local cluster reduced density matrices within subspaces constrained by local particle number and $\hat{S}^z$ conservation.
Further details regarding the implementation and matrix element construction are available in Refs. \cite{abraham_selected_2020, braunscheidel_generalization_2023}
For this paper, we have currently omitted this HOSVD step to avoid mixing local states that have different $\hat{S}^2$ eigenvalues. 
In the future, we will develop a fully spin-constrained HOSVD to allow applications for such systems.

\subsection{Heisenberg Hamiltonian}


As discussed in the introduction, Sec. \ref{sec:introduction}, multicenter TMCs play pivotal roles in different biological and mechanical processes because of their complex structures and spins.
Due to their diverse physical features, developing a single computational method that accurately captures their physics and low-energy spectrum remains a significant challenge.
The presence of multiple interacting spin centers results in multiconfigurational low-energy states, making conventional single-reference methods like perturbation theory and coupled-cluster inadequate for accurate description and even qualitatively inaccurate.

The Heisenberg-Dirac-van Vleck ($HDvV$) Hamiltonian represents the simplest phenomenological model for describing these magnetic interactions.
It can be derived from the Hubbard model at half-filling using quasi-degenerate perturbation theory.\footnote{For systems away from half-filling, alternative models such as the $tJ$ or double exchange Hamiltonians may be more appropriate.} 
For a system comprising two magnetic centers, labeled $a$ and $b$, the HDvV Hamiltonian takes the form:  
\begin{align}  
    \hat{H}^{HDvV} = -2J \hat{\vec S}_a \cdot\hat{\vec S}_b,  
\end{align}  
where $\hat{\vec S}_a = \hat{S}_a^x\vec x +\hat{S}_a^y\vec y +\hat{S}_a^z\vec z$ represents the spin operators for center $a$, and $J$ is the magnetic exchange coupling constant between the two centers, $a$ and $b$.
The sign and magnitude of $J$ determine the nature and strength of the spin coupling, with positive values corresponding to ferromagnetic (F) interactions and negative values indicating antiferromagnetic (AF) interactions.  

The low-energy spectrum of spin states in a two-center system can be qualitatively dictated by the $HDvV$ Hamiltonian and follows the Landé interval rule:  
\begin{align} \label{eq:lande_interval} 
    E(S) - E(S - 1) = -2SJ.  
\end{align}  
In the Hubbard model framework, the second-order contribution to the exchange coupling constant is quantified as \( J = \frac{-4t^2}{U} \), where \( t \) represents the electron hopping integral and $U$ is the on-site Coulomb repulsion. 
This so-called ``kinetic exchange" term describes how charge resonance or electron delocalization (characterized by 
$t$) contributes to the enhancement of antiferromagnetic interactions.
However, as highlighted in Ref. \citenum{faraday_discussions}, the zeroth-order ab initio Hamiltonian also has additional contributions beyond kinetic exchange, including non-local direct exchange interactions ($ K $).
While direct exchange generally favors high-spin states, kinetic exchange tends to stabilize low-spin configurations, making it challenging to determine both the sign and magnitude of $J$.

Although $J$ provides a useful criterion for interpreting magnetic interactions in multicenter complexes, the $HDvV$ Hamiltonian remains an approximate model of the true electronic structure.
Consequently, ab initio calculations are invaluable for not only do they enable the extraction of $J$ from computed energy differences between spin states, but they also assess the applicability of phenomenological Hamiltonians for a given system.
However, accurately computing these low-energy states remains a persistent challenge in computational chemistry, as multiple factors influence the relative energies of spin states in multicenter organometallic complexes, particularly those involving open-shell transition metal centers.~\cite{malrieuMagneticInteractionsMolecules2014}  

Beyond conventional exchange interactions, dynamic spin and charge polarization can significantly alter coupling strengths, further complicating the analysis.
Ligands play a crucial role in modulating magnetic interactions.
When metal ions are separated by a linear bridging ligand, direct metal-metal interactions can become negligible.
However, in multiply bridged dimers, variations in the bridging topology permit changes in metal-metal distances, which, in turn, affect the interaction strength.
This proximity effect can lead to deviations in antiferromagnetic coupling strengths from what is expected based solely on the chemical nature of the bridges.
Specifically, orbital mixing between metal centers and bridging ligands enhances the delocalization of unpaired electrons, thereby increasing magnetic coupling strength and introducing additional complexities in describing the interaction.

Due to these challenges, multireference methods such as CASSCF and CASPT2 are frequently employed to model exchange-coupled systems.
However, the high computational cost associated with these approaches constrains the size of the active space, often preventing results from reaching the level of accuracy necessary for meaningful quantitative comparisons with experimental data.
In many cases, dynamical correlation particularly interactions involving non-magnetic orbitals, plays a crucial role in determining the exchange coupling constant \( J \), typically enhancing antiferromagnetic interactions.  

Given these challenges, DMRG method has become the benchmark approach for computing exchange coupling constants in organometallic compounds \cite{pantazisMeetingChallengeMagnetic2019, szalayTensorProductMethod2015, baiardiDensityMatrixRenormalization2020, schollwockDensitymatrixRenormalizationGroup2011, woutersDensityMatrixRenormalization2014, eriksenGroundStateElectronic2020, olivares-amayaAbinitioDensityMatrix2015a, sharmaLowenergySpectrumIronsulfur2014, sharmaQuasidegeneratePerturbationTheory2016a, harrisInitioDensityMatrix2014}.
However, in cases where only the value of $J$ is required, more computationally efficient approaches that incorporate spin-flip methods have proven to be effective alternatives \cite{mayhallComputationalQuantumChemistry2014, mayhallComputationalQuantumChemistry2015a, mayhallSpinflipNonorthogonalConfiguration2014a, pokhilkoEffectiveHamiltoniansDerived2020,houckCombinedSpinFlipIP2019}.
In this paper, we explore the extent to which TPSCI can be used to compute exchange coupling constants in a range of TMC's (see Fig. \ref{fig:bimetallics}), 
comparing to DMRG results. 

\section{Computational Details}
In this paper, we explore the ability of TPSCI and DMRG to get comparable quantitative accuracy in terms of energy, $J$, and acceptable standard errors in energy extrapolation.
For this, we have chosen 7 TMCs (see Fig. \ref{fig:bimetallics}) and selected different sizes of active spaces (with different types of orbitals) to examine most of the important aspects of energy computation, extrapolation, and reliability of energies and $J$ values of the low-lying spin-states. 
We have employed the PySCF software~\cite{sun_recent_2020} for relevant geometry optimizations, Hartree-Fock calculations, and integral generations.
All TPSCI calculations are conducted using our open-source FermiCG software~\cite{mayhall_nicholas_fermicg_nodate}.
The necessary DMRG calculations are done using BLOCK2 software code~\cite{block2}.

\subsection{Linear Extrapolation of Energy and Standard Errors}\label{sec:extrap}
Converging to the exact eigenstates is challenging for both DMRG and TPSCI. 
However, when the convergence is sufficiently well-behaved, extrapolations to the converged limit can prove very useful.
DMRG extrapolation energy is achieved using the linear regression model for 
``variational energies'' vs ``discarded weight''
for different bond dimensions.
However, TPSCI extrapolation energy is achieved through linear extrapolation for ``variational energies'' vs ``PT2 correction''. 
The linear regression method aims to find the best-fitting straight line through a set of data points ($x_i, y_i)$ by minimizing the sum of squared residuals, $\text{SS}_{\text{resid}}$ which can be obtained by solving the following:
\begin{align}
    \min \text{SS}_{\text{resid}} = \min \sum_{i=1}^n (y_i-(mx_i +b))^2
\end{align}
where $m,b$ are the required slope and intercept parameters for the linear regression model.

The considered factors for this study that quantify the quality of the extrapolation are standard error of the y estimates ($\Delta s_y$), standard error of the y-intercept ($\Delta s_b$), and coefficient of determination ($R^2$), 
given as:
\begin{align}
    \Delta s_y =& \sqrt{\frac{\text{SS}_{\text{resid}}}{n - 2}} \\
    \Delta s_b =& \Delta s_y \sqrt{\frac{1}{n} + \frac{\bar{x}^2}{\sum (x_i - \bar{x})^2}} \\   
   R^2 =& \frac{\text{SS}_{\text{reg}}}{\text{SS}_{\text{tot}}},  
   \text{SS}_{\text{reg}} = \sum (\hat{y}_i - \bar{y})^2,
   \text{SS}_{\text{tot}} = \sum (y_i - \bar{y})^2,
\end{align}
   where $n$ is the number of points.
Of these quantities, we are most interested in $\Delta S_b$, as we care about the extrapolated energies ($y$ being the axis for energy). 
Extrapolated energy gaps are then used to extract the magnetic exchange coupling constants, $J$.
If we have two states and their corresponding standard errors, the standard error in $J$ can be calculated by the general error propagation formula which can be expressed as: \begin{align}\label{eq:propagation_error}
       \delta J = \frac{1}{2S} \sqrt{(\delta E_{S-1})^2 + (\delta E_S)^2},
   \end{align}
where $\delta E_S$ represents standard error corresponding to spin-state, $S$.

\subsection{Molecular Systems}
\begin{figure}[h]
    \centering
    \includegraphics[width=1\linewidth]{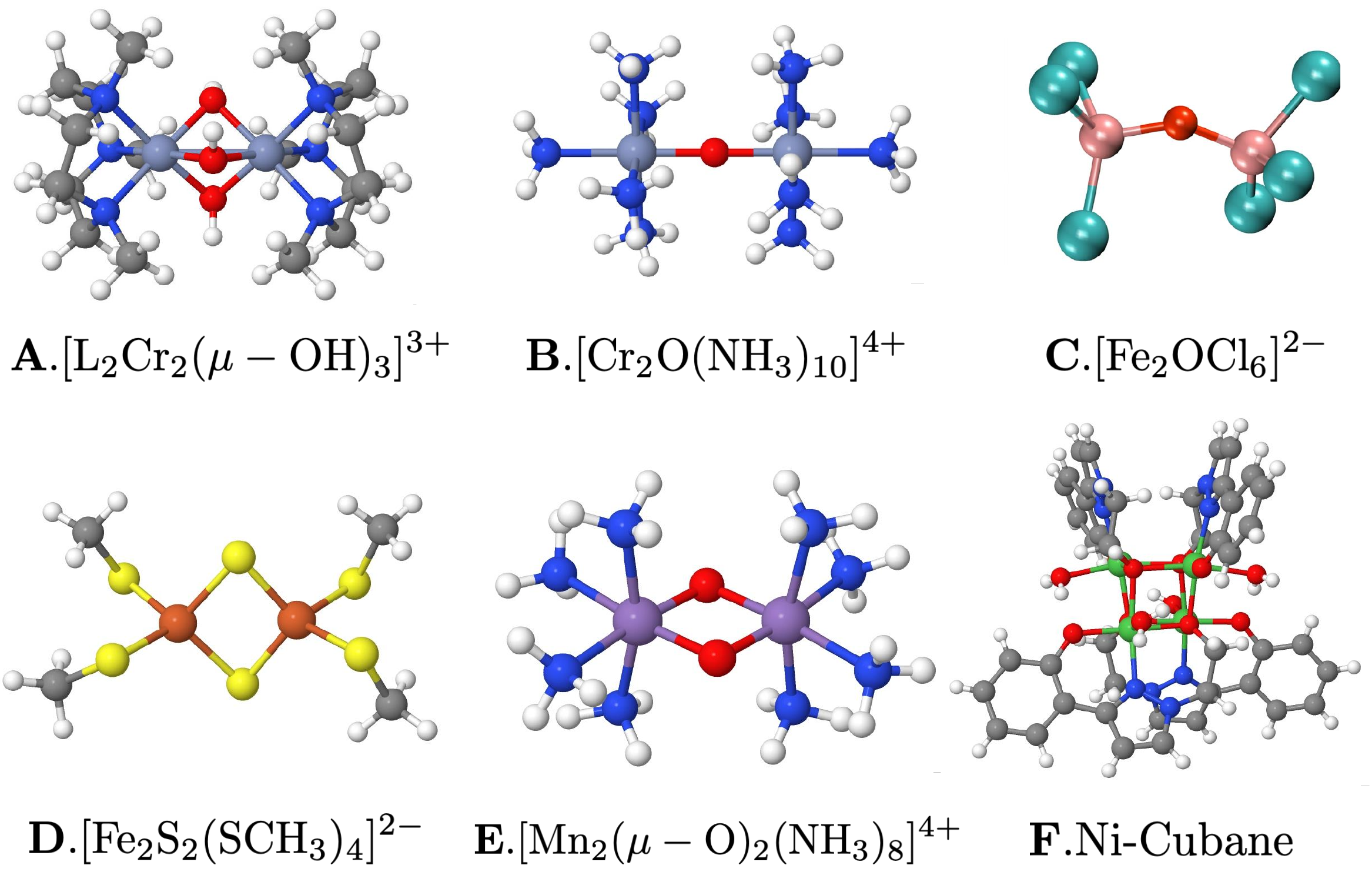}
    \caption{Transition Metal complex (TMCs) structures. \textbf{A}: $[\text{L}_2\text{Cr}_2(\mu-\text{OH})_3]^{3+}$, where L = $\text{N},\text{N}^{\prime}$,$\text{N}^{\prime \prime}$ $-$ trimethyl $-$ 1, 4, 7 $-$ triazacyclononane has six unpaired electrons with three on each Cr center. \textbf{B}: $[\text{Cr}_2\text{O}{(\text{NH}_3})_{10}]^{4+}$ also has six unpaired electrons with three on each metal center. \textbf{C}: $[\text{Fe}_2\text{OCl}_6]^{2-}$ has ten unpaired electrons with five on each iron center. \textbf{D}: $[\text{Fe}_2\text{S}_2{(\text{SCH}_3})_{4}]^{2-}$ also has ten unpaired electrons with five on each iron center in its oxidized form.
    \textbf{E}: $[\text{Mn}_2(\mu-\text{O})_2(\text{NH}_3)_8]^{4+}$ has six unpaired electrons with three electrons in each Mn center.
    \textbf{F}: Nickel-cubane complex has eight unpaired electrons with two electrons in each of the four nickel centers.}
    \label{fig:bimetallics}
\end{figure}

For this study, we have chosen five bimetallic compounds and a Ni-cubane complex, which are shown in Fig. \ref{fig:bimetallics}.
\subsection{Active Space Choice and clustering}\label{activespace}
In each of the systems studied, 
we follow a similar sequence of steps for defining the orbital active spaces that have strong overlap with metal d-orbitals and ligand orbitals, maintaining high-spin ROHF wavefunction with proper spin symmetry of the complex.
 We start by choosing atomic orbitals (AO) that have the desired character, and then we separately project the doubly occupied, singly occupied, and virtual orbital subspaces onto this AO subspace. 
 We then perform a singular value decomposition (SVD) on each subspace and retain the largest singular vectors from each of the ROHF subspaces to produce the same number of molecular orbitals as AO.
 This allows one to use chemical intuition to define the active spaces, 
 while at the same time, avoiding any orbital rotations that would destroy the ROHF reference.

The active space is divided into a suitable number of clusters which are then optimized in the RO-cMF procedure~\cite{bachhar_rocmf}. 
These optimized orbitals are used in the TPSCI calculations. 
Because DMRG depends on the ordering of the orbitals as well, we have used a couple different orderings in the calculations:``Fiedler" and ``GAOPT" algorithms. The ``Fiedler'' algorithm takes our initial cMF clusters orbital ordering as an initial guess to reorder them, while ``GAOPT" prefers Fiedler sorted orbitals as an initial guess to reduce the cost. More discussions on these can be found in Ref. ~\cite{block2}. 
We have used both procedures to see how this difference in orbital reordering results in absolute energies and energy gaps.

\subsection{Cluster basis formation}
For each cluster, a local many-body basis is constructed from the $M$ lowest energy states for each sector of Fock space which contained up to $N_I \pm \delta_{elec}$ number of electrons, by diagonalizing the cMF Hamiltonian.
For each active space, we solve for the lowest $M$ states in the lowest $M_s$ subspace, then generate higher $M_s$ vectors by applying directly $\hat{S}^{\pm}$.
This approach reduces the computational cost and ensures that every TPS in our basis can be properly spin-adapted to form eigenstates of $\hat{S}^2$. 

\section{Results and Discussion}
\subsection{Complex \textbf{A: }$[\text{L}_2\text{Cr(III)}_2(\mu-\text{OH})_3]^{3+}$}
Early experimental work on this di-chromium complex reported an exchange coupling constant, $J$, of $-66$ cm$^{-1}$ obtained through spectroscopic fitting procedures \cite{cr2_pantazis_experimental}.
More recently, Pantazis used DMRG-SCF with a (30e, 22o) active space, and reported a computed $J$ value of -23.9 cm$^{-1}$ \cite{pantazisMeetingChallengeMagnetic2019}.
In this paper, we use a larger (32e, 38o) active space for complex \textbf{A}, the details of which are provided in the Appendix, Sec. \ref{cr2_appendix}.

\subsubsection{Energy Convergence}
\begin{table*}
    \centering
    \begin{tabular}{c|cccc|ccc|ccc}
    \hline\hline &&TPSCI&&&&DMRG-FIEDLER&&&DMRG-GAOPT\\
         &var& $\Delta\text{E}_{\text{PT2}}$ & extrap & $\Delta S_b $& var &extrap &$\Delta S_b$ &var &extrap&$\Delta S_b$ \\\hline 
($S^2=12$) & $-0.186654$&	$-3.2$e-3&	$-0.191014$&	9.0e-5&$-0.205183$&	$-0.206137$&	4.5e-5&	$-0.205637$&	$-0.206323$&	1.9e-5\\
($S^2=6$)&$-0.187414$&	$-3.2$e-3	&$-0.191874$&	9.5e-5&	$-0.205772$&	$-0.20712$&	8.2e-5&	$-0.206282$&	$-0.207189$&	2.3e-5\\
($S^2=2$)&$-0.187904$&	$-3.2$e-3&$-0.192424$&9.9e-5&	$-0.206303$&	$-0.207888$&	2.0e-4&	$-0.206834$&	$-0.20772$&	9.9e-5\\
($S^2=0$)&$-0.188144$&	$-3.1$e-3	&$-0.192694$&	1.0e-4&	$-0.206637$&	$-0.208152$&	2.0e-4&	$-0.207186$&	$-0.207942$&	5.2e-5\\
         \hline\hline
\end{tabular}
    \caption{Energies for complex \textbf{A}. Absolute energies are shifted by $3344$ Hartrees. ``var" represents variational energy, ``$\Delta\text{E}_{\text{PT2}}$" represents ``PT2 correction" over variational energy, ``extrap" denotes extrapolated energy, and ``$\Delta S_{\text{b}}$" denotes standard error in y-intercept. DMRG variational energies are calculated at a bond dimension ($\chi$) of 1150.}
    \label{tbl:cr2_energy_tpsci_dmrg_pantazis}
\end{table*}

A series of TPSCI calculations were performed with different thresholds to yield a sequence of increasing variational space sizes.
As described in Sec. \ref{sec:extrap} (and as typically done in Selected CI applications), ~\cite{huron1973iterative,bender1969studies,whitten1969configuration,tubman2016deterministic,schriber2016communication,holmes2016heat,liu2016ici,chakraborty2018evolutionary,ohtsuka2017selected,levine2020casscf,kossoski2023seniority,kossoski2023state,kossoski2022hierarchy,chilkuri2021spin,loos2020performance,garniron2018selected,loos2018mountaineering,extrapolation_selected_ci}
we assume a linear relationship between the variational energies and PT2 energy corrections, allowing us to extrapolate to zero PT2 correction energy to obtain a final extrapolated TPSCI energy (approximation of exact diagonalization within the basis spanned by our cluster states).
Table \ref{tbl:cr2_energy_tpsci_dmrg_pantazis} shows the absolute energies and their corresponding standard errors in y-intercept ($\Delta S_b$) in energy extrapolation for TPSCI and DMRG for low-energy spin-states.

For both TPSCI and DMRG, energy decreases as $S^2$ decreases, consistent with an antiferromagnetic interaction.
If we consider the high-spin state ($S^2=12$), $\Delta S_b$ magnitudes are of the same order in all the approaches.
Reflecting the increased entanglement and decreased sparsity that arises from mixing in the charge resonance configurations necessary for achieving antiferromagnetic interactions, the $\Delta S_b$ increases as we move towards low-spin states.
While both orbital-ordering algorithms prove effective at achieving low-rank MPS solutions, there are subtle differences between the two approaches. 
The GAOPT algorithm results in DMRG extrapolations with slightly less extrapolation error than the FIEDLER algorithm.
Similarly, the variational energies are lower using GAOPT than FIEDLER, although extrapolated energies follow an opposite trend.

As mentioned earlier, the $M$ parameter defines a subspace of products of local states within which the TPSCI calculation is performed.
As such, a perfectly converged TPSCI calculation is only as accurate as the subspace it spans.
In order to check convergence with $M$, we perform multiple TPSCI calculations with different values of $M$ to ensure the results do not change significantly. 
Because the TPSCI results are converged only within the subspace spanned by $M$, DMRG is expected to yield lower absolute energies.
This is indeed the case, and Table \ref{tbl:cr2_energy_tpsci_dmrg_pantazis} illustrates the non-negligible impact of truncating the cluster state basis.

\subsubsection{Exchange Coupling constant}

\begin{table}[h]
    \centering
    \begin{tabular}{p{2cm}|p{1.2cm}p{1.2cm}p{1.2cm}p{1.2cm}}
    \hline\hline
         & $J$(0,1)& $J$(1,2)& $J$(2,3) & $J$(0,3)\\\hline
        & \multicolumn{4}{c}{TPSCI}  \\
        $M=100$ & $-25.4$ & $-25.7$ & $-26.6$&$-26.1$\\
        $M=200$ &  $-26.3$&$-26.9$&$-27.8$&$-27.3$\\
        & \multicolumn{4}{c}{TPSCI+PT2}  \\
        $M=100$ & $-26.7$ & $-27.3$ & $-28.3$&$-27.7$\\
        $M=200$ & $-28.1$ &$-28.8$&$-30.0$&$-29.3$\\
        & \multicolumn{4}{c}{TPSCI Extrapolated}  \\
        $M=100$ & $-28.0$ & $-28.7$ & $-30.0$&$-28.9$\\
        $M=200$ & $-29.3$ & $-30.3$ & $-31.3$&$-30.3$\\  
        & \multicolumn{4}{c}{DMRG-FIEDLER}  \\
        
        $Variational$ & $-36.65$ & $-29.13$& $-21.55$&$-26.60$\\
        $Extrapolated$ & $-28.91$ & $-42.13$ & $-35.97$&$-36.85$\\ 
        & \multicolumn{4}{c}{DMRG-GAOPT}  \\
        $Variational$ & $-38.64$ & $-30.29$ & $-23.56$&$-28.32$\\
        $Extrapolated$ & $-24.35$ & $-29.13$ & $-31.70$&$-29.61$\\ 

    \hline\hline
\end{tabular}
    \caption{Exchange coupling constants (cm$^{-1}$) for complex \textbf{A} with def2-SVP basis, and (32e, 38o) active space.
    $J$ refers to $H=-2J\hat{S}_1\cdot\hat{S}_2$.
    $J$(0,1) denotes which spin states are used to computed $J$ via Land\'e rule. ``TPSCI'' refers to the best variational energy obtained, using $\epsilon_{cipsi}=\num{2e-4}$. ``TPSCI+PT2'' is the best variational energy plus the state-specific PT2 correction.
    ``TPSCI Extrapolated"  refers to directly extrapolated $J$ values achieved through $J$ extrapolation route which involves linear extrapolation of ``TPSCI'' or ``TPSCI+PT2'' $J$ values vs contribution in $J$ from PT2 correction ($\Delta J$)
    $M$ is the maximum number of cluster states computed for each cluster Fock sector. The dimension of the sparse variational TPSCI subspace is 97357 for $M$=100 and 127493 for $M$=200, compared to the dimension of the full subspaces of 2.7 $\times 10^{14}$ and 6.1$ \times 10^{15}$ for $M=100$ and $M=200$, respectively.}
    \label{tbl:cr2_j_tpsci_pantazis}
    \end{table}

In principle, one could use any of the energy gaps from the computed energies of the spin-states (S=0,1,2, and 3) to get the exchange coupling constant ($J$) using  Land\'e interval rule (\ref{eq:lande_interval}) derived from two-site Heisenberg model.
If the Heisenberg model happened to be a complete description of the low-energy electronic structure, then the $J$ values extracted from these energy gaps would all be identical. 
For TPSCI, we have used all three types of energies (best variational energies, TPSCI+PT2 energies, and extrapolated energies) to compute $J$ values for comparative analysis.
Similarly, for DMRG, we have taken both the variational energies acquired for the largest bond dimension calculation (the largest possible on one of our compute nodes, which is also the same computational resources allocated for the TPSCI calculations) as well as the extrapolated results.
These results are shown in Table \ref{tbl:cr2_j_tpsci_pantazis}.

Examining the impact of energy extrapolation in TPSCI, we observe an increase in the $J$ values compared to TPSCI+PT2 by approximately 1 cm\(^{-1}\) in magnitude.
Furthermore, increasing $M$ from 100 to 200 results in only a 1 cm\(^{-1}\) increase in the $J$ value, despite the substantial expansion of the accessible Hilbert space dimension from \( 2.7 \times 10^{14} \) to \( 6.1 \times 10^{15} \).
TPSCI variational energies give an average $J$ value of $-27.3$ $cm^{-1}$ while extrapolated values give a $J$ value of $-30.3$ $cm^{-1}$.
While the extrapolation of TPSCI results yields a small, but non-zero, effect of the $J$ values, the overall Heisenberg behavior (all spin-gaps yielding similar $J$ values) remains almost unchanged between variational, PT2, and extrapolated values.

For comparison, we have also computed $J$ values using DMRG variational and extrapolated energies (with both Fiedler and GAOPT orderings).
Overall, we find variational DMRG results to be similar in magnitude to our TPSCI results when considering the $J$ values averaged over spin states.
However, the DMRG results exhibit significant non-Heisenberg behavior. 
It's interesting to note that the deviations from Land\'e intervals are \textsl{qualitatively} different between the Fiedler and GAOPT ordering algorithms. 
In future work, it would be worthwhile trying to better converge the DMRG results using multi-node parallelism to resolve which approach is more accurate for this system.

Overall, the $J$ values for both TPSCI and DMRG (when averaged over spin states) are in good agreement with the CASSCF(6e,10o)-NEVPT2 computed $J$ value of $-31.8$ $cm^{-1}$.~\cite{pantazisMeetingChallengeMagnetic2019}

\subsubsection{$J$ Extrapolation Route}
In Table \ref{tbl:cr2_j_tpsci_pantazis} above, we provided $J$ values computed by taking the difference of two extrapolated energies. 
Since each extrapolated energy has a linear regression error associated with it, the computation of energy differences must involve the proper error propagation following Eq. \ref{eq:propagation_error}.
However, for TPSCI, this is not the only way to extrapolate $J$ values. 
Since each TPSCI calculation produces multiple spin-states, we can choose to directly extrapolate the $J$ values instead of computing them from extrapolated energies.
In other words, instead of extrapolating the spin-states and then subtracting to obtain $J$, we can compute $J$ for each subspace size as well as the corresponding PT2 correction to $J$. This gives us data points to directly extrapolate $J$.
\begin{table}[h]
\begin{tabular}{c|cc|cc}
\hline\hline
& \multicolumn{2}{p{1.25in}|}{$J$ Extrapolation}  & \multicolumn{2}{p{1.25in}}{Energy Extrapolation}\\
& $J$ & $\Delta S_b(J) $ & $J$ & $\Delta S_b(J)$ \\\hline
$J$(0, 1) & $-29.3$  & $0.16$ & $-29.63$ & $15.49$ \\
$J$(1, 2) & $-30.3$ & $0.25$  & $-30.18$  & $7.53$ \\
$J$(2, 3) & $-31.3$   & $0.27$ & $-31.46$  & $4.78$\\
$J$(0, 3) & $-30.3$ & $0.22$  & $-30.42$& $2.47$\\
\hline\hline
\end{tabular}
\caption{Exchange coupling constants (cm$^{-1}$) for for complex \textbf{A} with def2-SVP basis, and (32e, 38o) active space in TPSCI. ``J Extrapolation'' refers to direct extrapolation of $J$ values, while ``Energy Extrapolation'' refers to $J$ values computed from  differences of extrapolated TPSCI energies. ``$\Delta S_b(J) $" refers to standard error in y-intercept for each case.}
    \label{tbl:cr2_j_tpsci_pantazis_error}
\end{table}
Table \ref{tbl:cr2_j_tpsci_pantazis_error} shows two different series of $J$ values and their corresponding standard errors, allowing us to compare the direct extrapolation of $J$ values to the computation of $J$ values from extrapolated energies.
From this data, it is evident that directly extrapolating the $J$ values results in much smaller linear regression errors compared to the computation of the $J$ values from differences of extrapolated energies. 
Nonetheless, it is interesting to observe that both approaches seem to yield very similar values for the extrapolated values, a result of the fact that the errors in the absolute energies are highly correlated, a feature of the data which is not captured by simple error propagation (Eq. \ref{eq:propagation_error}).

\begin{table}[h]
    \centering
    \begin{tabular}{p{1.5cm}|p{1.2cm}p{1.2cm}p{1.2cm}p{1.2cm}}
    \hline\hline
          & $J(0,1)$& $J(1,2)$&  $J(2,3)$& $J(0,3)$\\\hline
         &\multicolumn{4}{c}{DMRG-FIEDLER}  \\
  $J$ &$-28.91$ & $-42.13$ & $-35.97$&$-36.85$\\
       $\Delta S_b(J)$  &$31.27$ & $11.98$& $3.41$&$3.76$ \\
       
         &\multicolumn{4}{c}{DMRG-GAOPT}  \\
 $J$ & $-24.35$ & $-29.13$ & $-31.70$ & $-29.61$\\ 
      $\Delta S_b(J)$ &$12.30$ & $5.59$&$1.10$ &$0.34$ \\
        
    \hline\hline
\end{tabular}
    \caption{Exchange coupling constants (cm$^{-1}$) for for complex \textbf{A} with def2-SVP basis, and (32e, 38o) active space.
    $J$ refers to $H=-2J\hat{S}_1\cdot\hat{S}_2$.
    $J$(0,1) denotes which spin states are used to computed $J$ via Land\'e rule. ``DMRG-FIEDLER'' refers to the extrapolated energy obtained using the Fiedler algorithm orbital sorting. ``DMRG-GAOPT'' refers to the extrapolated energy obtained using genetic algorithm orbital sorting. ``$\Delta S_b(J) $" refers to standard error in $J$ using Eq. \ref{eq:propagation_error}}
\label{tbl:cr2_j_dmrg_pantazis}.
\end{table}


\begin{table*}
    \centering
    \begin{tabular}{c|c|cccc|ccc}
    \hline\hline Complex, &&&TPSCI&&&&DMRG-FIEDLER\\
      Active Space  & &var& $\Delta\text{E}_{\text{PT2}}$ & extrap & $\Delta S_b $& var &extrap &$\Delta S_b$  \\\hline 
&($S^2=12$)&$-0.788759$&	$-0.000392$  &$-0.789221$&	4.15e$-$6&      $-0.802820$&	$-0.802865$&	3.19e$-$6 \\
\textbf{B}. $[\text{Cr}_2\text{O}{(\text{NH}_3})_{10}]^{4+}$&($S^2=6$)&$-0.792200$&	$-0.0004646$  &$-0.792754$&	7.32e$-$6&	     $-0.806647$&	$-0.806741$&	3.02e$-$6 \\
(26e, 29o)&($S^2=2$)&$-0.794525$&	$-0.0004616$  &$-0.795084$&    8.08e$-$6&	     $-0.809181$&	$-0.809281$&	3.53e$-$6 \\
&($S^2=0$)&$-0.795590$&	$-0.0005279$	&$-0.796232$&	8.22e$-$6&	     $-0.810461$&	$-0.810527$&	7.50e$-$6 \\
         \hline

\textbf{C}. $[\text{Fe}_2\text{OCl}_6]^{2-}$&($S^2=30$)&$-0.919890$&	$-0.0001803$&$	-0.920152$&	4.55e$-$8&      $-0.932712$&	$-0.932716$&	6.55e$-$8 \\
(22e,29o) &($S^2=0$)&$-0.928397$&	$-0.0007976$  &$-0.928859$&	2.97e$-$7&	     $-0.943824$&	$-0.943850$&	7.99e$-$7 \\         \hline

\textbf{C}. $[\text{Fe}_2\text{OCl}_6]^{2-}$&($S^2=30$)&$-0.903919$&	$-0.0102683$  &$-0.916231$&	7.51e$-$5&      $-1.114295$&	$-1.122299$&	1.83e$-$4 \\
(42e, 49o) &($S^2=0$)&$-0.914318$&	$-0.0123898$  &$-0.929292$&	7.66e$-$5&	     $-1.122969$&	$-1.147895$&	2.99e$-$3 \\         \hline

\textbf{D}. $[\text{Fe}_2\text{S}_2{(\text{SCH}_3})_{4}]^{2-}$&($S^2=30$)&$-0.574453$&	$-0.0007976$  &$-0.575151$&	4.78e$-$6&      $-0.595483$&	$-0.595534$&	5.23e$-$6 \\
(22e,32o) &($S^2=0$)&$-0.580397$&	$-0.0005939$  &$-0.581329$&	6.93e$-$6&	     $-0.602859$&	$-0.603047$&	2.14e$-$5\\         \hline

\textbf{D}. $[\text{Fe}_2\text{S}_2{(\text{SCH}_3})_{4}]^{2-}$&($S^2=30$)&$-0.492067$&	$-$5.35e$-$6  &$-0.492069$&	3.89e$-$9&      $-0.492069$&	$-0.492069$&	$0.0$ \\
(46e, 28o) &($S^2=0$)&$-0.495499$&	$-$1.19e$-$5  &$-0.495502$&	2.39e$-$8&	     $-0.495502$&	$-0.495502$&	1.20e$-$9 \\         \hline


\textbf{E}. $\text{Mn}_2(\mu-\text{O})_2(\text{NH}_3)_8]^{4+}$&($S^2=12$)&$-0.888173$&	$-0.0105293$  &$-0.899770$&	6.79e$-$6&      $-0.972766$&	$-0.975612$&	2.59e$-$4 \\
(34e, 36o) &($S^2=0$)&$-0.891234$&	$-0.0117616$  &$-0.904018$&	5.36e$-$6&	     $-0.976853$&	$-0.981388$&	3.75e$-$4 \\

         \hline\hline
\end{tabular}
    \caption{``var" represents variational energies, ``$\Delta\text{E}_{\text{PT2}}$" represents PT2 correction energies over variational energies, ``extrap" denotes extrapolated energies, and ``$\Delta S_{\text{b}}$" denotes standard error in y intercept. Absolute energies of complex \textbf{B}, \textbf{C}, \textbf{D}, \textbf{E} are shifted by $2721$, $5356$, $5067$, $6034$ Hartrees, respectively.  DMRG variational energies are calculated at MPS bond dimension ($\chi$) of 1600, 1000, 1500, and 1000 for complex \textbf{B}, \textbf{C}, \textbf{D}, \textbf{E}, respectively. }
    \label{tbl:cr2_energy_tpsci_dmrg_all}
\end{table*}
An analogous $J$-extrapolation approach was not immediately obvious for the DMRG results, simply because of the state-specific nature of the MPS optimization. However, one could imagine a multistate MPS approach, but we have not explored this beyond preliminary tests which did not appear promising.
In Table \ref{tbl:cr2_j_dmrg_pantazis}, we compute the extrapolation errors of the DMRG results, and observe that both orbital ordering algorithms yield significantly higher linear regression errors for the low spin states.
Because of the higher linear regression errors in the low-spin state $J$ values, it's difficult to conclude whether complex \textbf{A} is truly non-Heisenberg and the differences from TPSCI are qualitative, or if the DMRG results are still too far from converged.
More extensive (i.e., multinode parallelized) computations will be needed to resolve this. 

For the rest of the complexes,  \textbf{B}, \textbf{C}, \textbf{D}, \textbf{E}, and \textbf{F}, we report DMRG calculations using the ``Fiedler" sorted orbitals, unless ``GAOPT" is explicitly mentioned.

\subsection{Complex \textbf{B: }$[\text{Cr}_2\text{O}{(\text{NH}_3})_{10}]^{4+}$}


The dinuclear complex $[$Cr$_2$O(NH$_3)_{10}]^{4+}$ has a long history of being been scrutinized computationally, with a variety of methods from DFT to DMRG, and IC-MRCI methods being used to explore its magnetic properties.~\cite{REN2005159,harris_ab_2014} 
Morokuma~\cite{harris_ab_2014} advocated the neglect of the complete set of d-double-shell orbitals, as Cr(III) has two unoccupied $d$ orbitals.
Although non-magnetic orbitals and their double shells typically exert minimal influence on magnetic coupling, incorporation of the unoccupied Cr($d_{z^2}$) orbital resulted in a notably enhanced active space.
This enhancement ultimately leads to stronger magnetic coupling in the case of Cr(III).
Taking all these studies into account, we proceed to choose our active space, taking care to include enough orbitals to allow all possible interactions among the nominally $3d$, $4d$, and $4s$ Cr orbitals and the $2p$, $3s$, and $3p$ bridging O orbitals.
The detailed description of the selected active space, (26e, 29o), can be found in Appendix section \ref{cr2_2_appendix}.

\subsubsection{Energy Convergence}

Since each Cr(III) center has three unpaired electrons, the low-energy spectrum includes four spin states with $S^2$ values of $\{0, 2, 6, 12\}$.  
TPSCI calculations are performed for both $M$=200 and 250, and DMRG variational energies are achieved at the MPS bond dimension ($\chi$) of 1600.
TPSCI energy levels for $M$=250 and DMRG extrapolation plots are shown in Fig. \ref{fig:Cr_complex} (see Appendix). 
Compared to TPSCI, DMRG yields lower energies while exhibiting similar standard errors in the y-intercept ($\Delta S_b(J) $) (see Table \ref{tbl:cr2_energy_tpsci_dmrg_all}). 
As discussed earlier, the truncation of TPSCI to $M=250$ in the cluster basis prevents TPSCI absolute energies from converging closer to the FCI limit.
\subsubsection{Exchange Coupling Constants}
\begin{table}[h]
    \centering
    \begin{tabular}{c|cccc}
    \hline\hline
         & $J$(0,1)& $J$(1,2)& $J$(2,3)&$J$(0,3)\\\hline
        & \multicolumn{4}{c}{TPSCI}  \\
        $M=200$ & $-117.6$ & $-124.1$ & $-123.9$&$-122.9$\\
        $M=250$ & $-116.9$ & $-127.6$ & $-125.9$&$-124.9$\\
        & \multicolumn{4}{c}{TPSCI+PT2}  \\
        
        $M=200$ & $-121.3$ & $-124.0$ & $-125.5$&$-124.3$\\
        $M=250$ & $-124.2$ & $-127.4$ & $-128.5$&$-127.4$\\
        & \multicolumn{4}{c}{TPSCI Extrapolated }  \\
        
        $M=200$ & $-122.4$   & $-123.8$  & $-125.8$ &$-124.7$\\
        & (0.06)  & (0.01) & (0.03)&(0.02)\\
        $M=250$ & $-126.0$ & $-127.9$ & $-129.2$ &$-128.2$\\ 
        & (0.21)& (0.27)& (0.13)&(0.06)\\\hline
        
        & \multicolumn{4}{c}{DMRG-Fiedler}  \\
        $Variational$ & $-140.4$&$-139.0$&$-140.0$&$-139.7$\\
        $Extrapolated$ & $-136.7 $ &$-139.4$&$-141.7$&$-140.1$\\
        &(0.91)& (0.26)&(0.16)&(0.15)\\
    \hline\hline
\end{tabular}
    \caption{Exchange coupling constants (cm$^{-1}$) for for complex \textbf{B} with def2-SVP basis, and (26e, 29o) active space.
    $J$ refers to $H=-2J\hat{S}_1\cdot\hat{S}_2$.
    $J$(0,1) denotes which spin states are used to computed $J$ via Land\'e rule.
    ``TPSCI'', ``TPSCI+PT2'', ``TPSCI-Extrapolated" and $M$ have been defined in Table \ref{tbl:cr2_j_tpsci_pantazis}. Standard errors ($\Delta S_b(J)$) corresponding to extrapolated $J$ values are given in the parentheses. $\Delta S_b(J)$ refers to the standard error in direct $J$ extrapolation for TPSCI and the standard error in $J$ for DMRG derived using Eq. \ref{eq:propagation_error}.
    The dimension of the sparse variational TPSCI subspace is 13.5K and 32K for $M$=200 and 250, respectively.}
    \label{tbl:cr2_j_tpsci_morokuma}
\end{table}

As observed in Table \ref{tbl:cr2_j_tpsci_morokuma}, the exchange coupling constants ($J$) computed using TPSCI exhibit a noticeable dependence on the number of states in cluster basis ($M$). 
Increasing $M$ from 200 to 250 leads to a systematic increase in the magnitude of $J$, suggesting that a larger $M$ could capture more electron correlation and improve the accuracy of energy gaps.
The trend is consistent across all $J$ values, indicating that TPSCI could still systematically go beyond the reported $J$ values with increasing $M$.
The PT2 corrected and extrapolated energy gaps exhibit a similar trend while $J$ values tend to follow the Heisenberg spectrum more accurately, suggesting that further increasing $M$ would likely yield even more accurate estimates.

Comparing TPSCI with DMRG-Fiedler results, we observe that DMRG variational calculations produce significantly more negative exchange coupling constants, meaning stronger antiferromagnetic interactions.
The extrapolated DMRG values are more or less the same as the variational ones, indicating convergence of the DMRG results with bond dimension ($\chi=1600$).
Table \ref{tbl:cr2_j_tpsci_morokuma} further compares the standard errors in $J$ for TPSCI and DMRG methods.
Notably, the standard errors ($\Delta S_b(J)$) are small in magnitude in both methods, 
indicating that the extrapolations are expected to be reliable.

Despite the differences in absolute values, both methods capture the same qualitative trend.
Overall, the results confirm that while TPSCI provides a reliable estimate of exchange couplings, its accuracy can be strongly influenced by the choice of $M$.
The observed convergence behavior suggests that further refinement, either by increasing $M$ and improving cluster basis or incorporating improved extrapolation techniques, could enhance agreement with DMRG results.

\subsection{Complex \textbf{C: }$[\text{Fe}_2\text{OCl}_6]^{2-}$}
Previously, the exchange coupling constant between the two Fe centers in \([ \text{Fe}_2 \text{OCl}_6 ]^{2-}\)  has been examined using various theoretical approaches, including unrestricted Hartree-Fock (UHF),~\cite{uhf_fe} DFT,~\cite{dft_fe1,dft_fe2} internally contracted multireference configuration interaction (IC-MRCI),~\cite{ic_mrci} Spin-Flip methods~\cite{mayhallComputationalQuantumChemistry2014}, and DMRG methods~\cite{harris_ab_2014,block2}.
These studies report different $J$ values for different active spaces, making it difficult to compare different approaches for computing exchange coupling constants. 
In this article, we have chosen two different active spaces: 
a smaller (22e, 29o) active space which allows tight convergence with DMRG for benchmarking, 
as well as a larger (42e, 49o) active space which contains a more complete set of orbitals, 
which is described in Section \ref{fe2_appendix}.


\subsubsection{Energy Convergence}
For the smaller (22e, 29o) active space, both TPSCI and DMRG achieve good convergence with small standard errors in the extrapolated energies (see Table \ref{tbl:cr2_energy_tpsci_dmrg_all}).
For example, in the high-spin state ($S^2 = 30$), the DMRG extrapolated energy has a very small $\Delta S_b$ of $6.55 \times 10^{-8}$ $E_h$.
Similarly, for the low-spin state ($S^2 = 0$), the extrapolated energy comes with a slightly larger but still very small $\Delta S_b$, $7.99 \times 10^{-7}$ $E_h$.
These results indicate that DMRG is able to capture all relevant correlations in this active space while maintaining computational efficiency. 
TPSCI energies are also well converged  as it has small $\Delta S_b$ and PT2 correction to the variational energy, $\Delta\text{E}_{\text{PT2}}$.
As we have seen for other TM complexes, TPSCI absolute energies are a bit higher than DMRG energies.

When moving to the larger (42e, 49o) active space, 
we notice a significantly increased impact of the $M$ cluster basis truncation in TPSCI results. 
Because the clusters centered on the metal centers have an increased number of orbitals (11 on each Fe center) (see Fig. \ref{fig:fe2ocl6_complex}), 
a similar cluster state truncation projects out a much larger percentage of the overall Hilbert space. 
In fact, the singlet ground state energy, actually increases in absolute energy when increasing the active space size. 
This indicates that $M$ cluster states chosen for this system is far from the optimal set. 
In contrast, the DMRG absolute energies are significantly lowered upon increasing active space size, albeit with a larger standard error of $1.83 \times 10^{-4}$ $E_h$ and $2.99 \times 10^{-3}$ $E_h$ for $S^2$=30 and 0 spin-states, respectively. 

\subsubsection{Exchange coupling constants}
Table \ref{tbl:fe2_j_tpsci} presents exchange coupling constants ($J$ in cm$^{-1}$) for Fe$_2$ anionic complex (complex \textbf{C}), calculated using 6-31g* basis set for both active spaces: (22e, 29o) and (42e, 49o).
The exchange coupling constants, $J$ computed between consecutive spin states (0-1 to 4-5), are determined using the Landé rule for TPSCI, TPSCI+PT2. We have used TPSCI $J$ values and difference between TPSCI+PT2 and TPSCI $J$ values to do the linear interpolation for $J$ (J extrapolation route). 
Across both active spaces, the $J$ values generally become more negative when moving from TPSCI to TPSCI+PT2 and subsequently to TPSCI extrapolated results, indicating enhanced inclusion of electron correlation effects, and importance of extrapolation,
but the larger active space has a significantly larger effect.

\begin{table}
  \centering
\begin{tabular}{c|ccccc|c}\hline \hline
$M$   & J(0,1) & J(1,2)  & J(2,3) & J(3,4) & J(4,5) & J(0,5)  \\
\hline
& \multicolumn{5}{c|}{(22e, 29o)} \\\hline
& \multicolumn{5}{c|}{TPSCI} \\
200	&$-61.11$	&$-61.17$	&$-61.41$	&$-61.70$	&$-61.99$ &$-61.63$ \\
250&$-61.78$	&$-61.76$&	$-62.05$&	$-62.43$	&$-62.83$&	$-62.35$\\
& \multicolumn{5}{c|}{TPSCI+PT2}   \\
200&	$-61.55$	&$-61.66$	&$-61.87$&	$-62.11$&	$-62.34$ & $-62.04$\\
250&$-62.78$&	$-62.91$&	$-63.18$&	$-63.51$&	$-63.85$&	$-63.43$\\
& \multicolumn{5}{c|}{TPSCI Extrapolated}   \\
200	&$-61.63$&	$-61.75$&	$-61.96$&	$-62.18$&	$-62.42$ & $-62.13$\\
 &(2e$-$3) & (2e$-$3)& (9e$-$4) & (2e$-$2) & (2e$-$3) & (5e$-$3)\\
250	&$-63.01$&	$-63.12$&	$-63.35$&	$-63.59$&	$-64.13$ & $-63.83$\\ 
 & (1e$-$2)& (7e$-$3)&(5e$-$3)&(3e$-$2) & (2e$-$3) &(2e$-$3) \\
\hline
& \multicolumn{5}{c|}{DMRG Variational} \\
& -- &  -- &  -- & -- &-- &$-81.29$\\
& \multicolumn{5}{c|}{DMRG Extrapolated} \\
& -- &  -- &  -- & -- &-- &$-81.45$ \\
&    &     &     &    & &(6e-3)\\\hline
& \multicolumn{5}{c|}{(42e, 49o)} \\\hline
& \multicolumn{5}{c|}{TPSCI}   \\
100 & $-77.08$  & $-77.72$ & $-78.76$ & $-80.18$   & $-81.77$ & $-79.89$ \\
150 & $-72.12$ & $-74.44$ & $-75.75$& $-76.37$   & $-77.50$&  $-76.08$ \\

& \multicolumn{5}{c|}{TPSCI+PT2}   \\
100 & $-89.53$& $-90.13$ & $-91.09$ & $-92.57 $  &$ -94.40$ & $-92.36$ \\
150 & $-88.58$ & $-89.45$& $-90.39$  & $-91.73$   & $-93.47$ & $-91.53$ \\

& \multicolumn{5}{c|}{TPSCI Extrapolated}   \\
100& $-90.89$&$-91.45$&$-93.11$&$-94.69$&$-97.63$&$-94.68$\\
&(0.20)&(0.31)&(0.15)&(0.19)&(0.13)&(0.08)\\
150&$-92.14$&$-92.16$&$-93.61$&$-95.56$&$-98.36$&$-95.46$\\
& (0.79)& (0.24)&(0.12)&(0.09)&(0.29)&(0.11)\\
\hline
& \multicolumn{5}{c|}{DMRG Variational} \\
& -- &  -- &  -- & -- & --&$-63.46$\\
& \multicolumn{5}{c|}{DMRG Extrapolated} \\
& -- &  -- &  -- & -- &-- &$-187.25$ \\
&    &     &     &    & &(21.88)\\\hline\hline

\end{tabular}
 \caption{Exchange coupling constants (cm$^{-1}$) for complex \textbf{C} with 6-31g* basis.
    $J$ refers to $H=-2J\hat{S}_1\cdot\hat{S}_2$.
    $J$(0,1) denotes which spin states are used to computed $J$ via Land\'e rule.
    ``TPSCI'', ``TPSCI+PT2'', ``TPSCI Extrapolated", and ``$M$'' have been defined in Table \ref{tbl:cr2_j_tpsci_pantazis}. ``$\Delta S_b$'' is defined in Table \ref{tbl:cr2_j_tpsci_morokuma}.
    The dimension of the sparse variational TPSCI subspace is  161.6K for $M$=200 and  345K for $M$=250 for (22e, 29o) active space.
    However, the dimension of the TPSCI subspace is  202K for $M$=100 and  141K for $M$=150 for (42e, 49o) active space.}
    \label{tbl:fe2_j_tpsci}
\end{table}

For the smaller active space (22e, 29o), TPSCI provides extrapolated $J$ values for consecutive spin states (0$-$1, 1$-$2, etc.), which range from $-63.01$ cm$^{-1}$ to $-64.13$ cm$^{-1}$.
The standard error ($\Delta S_b(J)$) values are very small, typically on the order of $10^{-2}$ cm$^{-1}$ or smaller, indicating excellent convergence within the space spanned by our $M$ truncation.
DMRG results for the same active space provide an average $J(0,5)$ of $-$81.29 cm$^{-1}$ and $-$81.45 cm$^{-1}$ from variational and extrapolated energy gaps, respectively.
These values are significantly more negative than TPSCI results,
indicating that the $M$ truncation in TPSCI is having a significant impact on the results. 
Worse still, the difference in $J$ value between $M=200$ and $M=250$ is negligible, 
indicating that simply increasing $M$ is not a good enough test for checking convergence in the number of cluster states. 

For the larger active space (42e, 49o), both TPSCI and DMRG provide more negative exchange coupling constants due to the inclusion of additional orbitals and electrons, which enhance correlation effects. 
For example, TPSCI extrapolated $J$ values range from $-$92.14 cm$^{-1}$ to $-$98.36 cm$^{-1}$, while those for the (22e, 29o) active space are noticeably smaller in magnitude.
However, when considering the larger cluster sizes in the (42e, 49o) active space, the larger dimension of the local cluster Hilbert space required us to use a looser $\epsilon_{CIPSI}$ threshold for the larger $M$ calculation. 
As a result, the TPSCI variational space dimension for $M$=100 (202K) is actually larger than that for $M$=150 (141K).

DMRG results in this active space give us an unreliable $J$ value, with an extrapolated $J$ of -187.25 cm$^{-1}$ with a large standard error of $\Delta S_b(J) = 21.88$ cm$^{-1}$.
Alternatively, using a different orbital reordering (GAOPT), DMRG yields $J_{GAOPT} = -144.69$ cm$^{-1}$ with a smaller standard error of $\Delta S_b(J) = 7.69$ cm$^{-1}$. 
For comparison, the experimentally derived $J$ value for this di-iron complex is $-117$ cm$^{-1}$.~\cite{dft_fe1}
The lack of convergence with DMRG is a consequence of the fact that due to the large active space size, we were limited to a bond dimension of $\chi=1000$, as larger values demanded too much memory for a single compute node.

\subsection{Complex \textbf{D: } $[\text{Fe}_2\text{S}_2{(\text{SCH}_3})_{4}]^{2-}$ }
Iron-sulfur clusters are molecular assemblies consisting of multiple (usually two to eight) iron (Fe) atoms interconnected by sulfur (S) ligands.
These clusters play crucial roles in fundamental biological processes such as nitrogen fixation, photosynthesis, and respiration.
Their functions encompass a wide array of activities, including redox chemistry, electron transfer, and even oxygen sensing.\cite{fe2s2_1,fe2s2_2,fe2s2_3}
The electronic structures of these clusters are crucial, as they host multiple low-lying states with distinct electronic and magnetic properties.
A detailed understanding of their electronic configurations is essential for elucidating the fundamental chemical mechanisms that govern these biologically significant processes.

Complex \textbf{D}, $[$Fe$_2$S$_2($SCH$_3)_4]^{2-}$ can be a challenging system for active space methods, aiming to capture a significant portion of its correlation energy and accurately predict both qualitative and quantitative properties.
It is important to choose a large active space that considers the configurations involving Fe $d$ electrons with contributions from excitations of electrons from bridging S ligands or even the S ligand orbitals.
In order to observe the impact of active space size, we again have chosen two different active spaces (described in detail in Appendix section \ref{fe2s2_appendix}): (46e, 28o) and (22e, 32o), where the larger 32 orbital active space includes additional virtual orbitals to better capture electron correlation effects, particularly from Fe's $4d$ and S's $4p$ orbitals.

\subsubsection{Energy Convergence}
Table \ref{tbl:cr2_energy_tpsci_dmrg_all} shows energies and standard errors for the complex \textbf{D}, focusing on the impact of different active spaces and spin states on computed energies.
For both active spaces, TPSCI calculations are performed for all $S^2=\{0,2,6,12,30\}$ states (see Appendix \ref{fe2s2_appendix}), while DMRG calculations are performed for only high-spin ($S^2 = 30$) and low-spin ($S^2 = 0$) states.

The smaller (46e, 28o) active space contains a minimal active space for each Fe cluster, i.e. ROHF subspace of (5e, 5o) while having 6 sulfur ligand clusters comprised of the 3p occupied orbitals. More detailed description can be found in Appendix \ref{fe2s2_appendix}.
The energies for the (46e, 28o) active space are essentially identical for both TPSCI and DMRG methods, as shown in Table \ref{tbl:cr2_energy_tpsci_dmrg_all}.
This is a result of the fact that no cluster state truncation is necessary for TPSCI, because the metal clusters contain only 5 orbitals for the smaller active space, 
again emphasizing the limiting role that cluster state truncation has on the final results. 


For the larger (22e, 32o) active space, the effect of the $M$ truncation re-emerges.
The DMRG extrapolated absolute energies are around 20 mH lower for DMRG than for TPSCI.


\subsubsection{Exchange coupling constants}
Table \ref{tbl:fe2s2_j_tpsci} presents the $J$ values and standard errors, $\Delta S_b$ for complex \textbf{D}.
As we have identical absolute energies for the (46e, 28o) active space, both TPSCI and DMRG calculations yield the same $J$ values. 

\begin{table}[ht]
\begin{tabular}{c|ccccc|c}\hline \hline
Active space& J(0,1) & J(1,2)  & J(2,3) & J(3,4) & J(4,5)&J(0.5) \\
\hline
& \multicolumn{5}{c|}{TPSCI}& \\
(46e, 28o)&$-25.04$&	$-25.24$&	$-25.08	$&$-25.07$&	$-24.97$&$-25.06$\\
(22e, 32o)&$-43.34$&	$-43.30$&	$-43.39$	&$-43.52$&	$-43.61$&$-43.48$\\

&\multicolumn{5}{c|}{TPSCI+PT2} &  \\
(46e, 28o)&$-25.22$&	$-25.24	$&$-25.17$&	$-25.11$&	$-25.00$&$-25.11$\\
(22e, 32o)&	$-44.82$&	$-44.86$&	$-44.93$&	$-45.01$&	$-45.04$&$-44.97$\\

& \multicolumn{5}{c|}{TPSCI Extrapolated} &  \\
(46e, 28o)&$-25.26$&	$-25.24$	&$-25.19$	&$-25.11$&$-25.26$&$-25.11$\\
&(5e$-$4)&(2e$-$4)&(1e$-$3)&(7e$-$4)& (8e$-$4)&(5e$-$4)\\

(22e, 32o)&$-45.09$&	$-45.22$&	$-45.21$&	$-45.33$	&$-45.30$&$-45.48$\\
&(2e$-$2)&(4e$-$2)&(1e$-$2)&(3e$-$3)&(2e$-$2)&(0.05)\\
\hline 
& \multicolumn{5}{c|}{DMRG Variational} \\
(46e, 28o)& -- &  -- &  -- & -- &--&$-25.12$ \\
(22e, 32o)& -- &  -- &  -- & -- &-- &$-53.96$\\
& \multicolumn{5}{c|}{DMRG Extrapolated} \\
(46e, 28o)& -- &  -- &  -- & -- & --&$-25.12$ \\
&    &     &     &    & &(9e$-$6)\\
(22e, 32o)& -- &  -- &  -- & -- & --&$-54.97$ \\
&    &     &     &    & &(0.16)\\\hline\hline

\end{tabular}
 \caption{Exchange coupling constants (cm$^{-1}$) for complex \textbf{D} with  basis. ``J(0,1)", ``TPSCI'', ``TPSCI+PT2'', and ``TPSCI Extrapolated'' have already been defined in Table \ref{tbl:cr2_j_tpsci_pantazis}. ``$\Delta S_b(J)$'' values for extrapolated $J$ are given in parenthesis.
    The sparse variational TPSCI subspace dimension is  192.8K for $M$=200 for (22e, 32o) active space and 164.4K for (46e, 28o). 
    }
    \label{tbl:fe2s2_j_tpsci}
\end{table}

For the larger (22e, 32o) active space, $J$ values calculated using TPSCI and DMRG show notable differences, both in magnitude and associated standard errors ($\Delta S_b(J)$).
TPSCI provides extrapolated $J$ values for each of the adjacent spin state differences ($S=1,0$ to $S=5,4$), with values ranging from $-$45.09 cm$^{-1}$ to $-$45.33 cm$^{-1}$, each with relatively small $\Delta S_b(J)$.
Since all energy gaps yield essentially the same $J$ value, 
the TPSCI results indicate that the system is well-described by the Heisenberg Hamiltonian.

DMRG, on the other hand, yields a variational $J$ value of $-$53.96 cm$^{-1}$ and extrapolated $J$ value of $-$54.97 cm$^{-1}$ for the (S=5,0) spin-state energy gap.
Whereas DMRG and TPSCI provided near identical results for the smaller 28o active space, the methods differ non-trivially for the larger 32o active space, with DMRG predicting more strongly antiferromagnetic interactions.

\subsection{Complex \textbf{E: }$\text{Mn}_2(\mu-\text{O})_2(\text{NH}_3)_8]^{4+}$}

Mn-oxo dimers are center to many catalytic processes and redox reactions and represents as a important system for carrying out a detailed investigation using TPSCI and DMRG.
A recent study~\cite{stein2019orbital} predicted the coupling constant ($J_{BS}$) using broken-symmetry density functional theory (BS-DFT) with the TPSSh density functional is $-$115 $cm^{-1}$.
This indicates an antiferromagnetic coupling between the Mn centers.
The molecular orbitals and active space choice for this study are described in Appendix section \ref{mn2_appendix}.


\subsubsection{Energy Convergence}

The presence of three unpaired electrons at both Mn centers lead to four low-lying spin states, $S^2 = \{0, 2, 6, 12\}$.  
TPSCI calculations are performed for both $M$=100 and 150, and DMRG variational energies are achieved for the MPS bond dimension $\chi=1000$.
TPSCI energy levels for $M$=150 are shown in Fig. \ref{fig:Mn_complex} (I).

From Table \ref{tbl:cr2_energy_tpsci_dmrg_all}, we can see that the TPSCI extrapolated energy is approximately 2500 $cm^{-1}$ lower than the variational energy, while for DMRG the difference is less (approximately 1000 $cm^{-1}$).
However, the standard error range for DMRG extrapolated energies is larger (approximately $3 \times 10^{-4}$ $E_h$) than in TPSCI ($2 \times 10^{-6}$ $E_h$ to $9 \times 10^{-6}$ $E_h$). A plot of the convergence is given in the Appendix in Fig. \ref{fig:Mn_complex}. 

\subsubsection{Magnetic exchange coupling constant}

\begin{table}[]
\begin{tabular}{p{1.0cm}|p{1.5cm}p{1.5cm}p{1.5cm}|p{1.5cm}}
\hline \hline
$M$   & \hspace{0.15cm}J(0,1) & \hspace{0.15cm}J(1,2)     & \hspace{0.15cm}J(2,3) & \hspace{0.15cm}J(0,3)\\ \hline 
 &          & TPSCI        &            & \\
100 & $-53.67$   & $-53.53$& $-55.60$  &$-54.59$ \\
150 & $-54.69$    &$ -55.15$& $-56.98$  & $-55.99$\\
&           \multicolumn{3}{c|}    {TPSCI+PT2}       \\
100 & $-71.26$  & $-70.89$& $-73.17$  & $-72.09$\\
150 & $-77.71 $ & $-77.08$& $-79.76$&$-78.53$\\
&       \multicolumn{3}{c|}    {TPSCI Extrapolated} \\
100 & $-73.26$    & $-73.03$ & $-75.96$ &$-75.23$\\
& \hspace{0.12cm}(0.06)   & \hspace{0.12cm}(0.03)& \hspace{0.12cm}(0.06)&\hspace{0.12cm}(0.24)\\
150& $-79.94$ & $-79.40$  & $-83.92 $& $-83.98$\\ 
& \hspace{0.12cm}(0.13)& \hspace{0.12cm}(0.01)& \hspace{0.12cm}(0.07) &\hspace{0.12cm}(0.31)\\

\hline
& \multicolumn{3}{c|}{DMRG Variational} \\
& \hspace{0.25cm} -- & \hspace{0.25cm} -- &  \hspace{0.25cm} --  &$-74.99$\\
& \multicolumn{3}{c|}{DMRG Extrapolated} \\
&\hspace{0.25cm} -- & \hspace{0.25cm} -- & \hspace{0.25cm} --  &$-105.64$ \\
&    &     &   &\hspace{0.12cm}(8.34)\\
 \hline \hline                        
\end{tabular}
 \caption{Exchange coupling constants (cm$^{-1}$) for the Mn$_2$ complex (\textbf{E}) with  def2-SVP basis. ``J(0,1)", ``TPSCI'', ``TPSCI+PT2'', ``TPSCI Extrapolated", and ``$M$" have already been defined in Table \ref{tbl:cr2_j_tpsci_pantazis}. Standard errors ($\Delta S_b(J)$) in extrapolated J computation are given in the parentheses.
    The sparse variational TPSCI subspace dimension is  128K for $M$=100 and 117K for $M$=150 for (34e, 36o) active space.}
    \label{tbl:Mn2_j_tpsci}
\end{table}

Table \ref{tbl:Mn2_j_tpsci} presents the exchange coupling constants, $J$, for complex \textbf{E} calculated using TPSCI and DMRG.
For $M$ = 100 and $M$ = 150, the sparse variational TPSCI subspace dimensions are 128K and 117K, respectively.
The observed trends suggest that TPSCI provides a reliable but slightly underestimated description of the exchange interactions, while TPSCI+PT2 and extrapolated results offer a more complete treatment.
The relatively small difference between TPSCI+PT2 and extrapolated values indicates that the perturbative correction effectively captures the missing inter-cluster correlations.


Comparing the $J$ values obtained using the various spin-state energy gaps, we see that modest non-Heisenberg behavior is observed, with the $S=2,3$ gap consistently yielding a slightly larger value. 
The relatively small values of $\Delta S_b(J)$ in TPSCI suggest that the computed $J$-values are well-converged within the selected truncation scheme, indicating a reasonable level of accuracy in capturing strong correlation effects within the (34e, 36o) active space.  

For the DMRG calculations, we have considered two types of orbital reordering: ``Fiedler" and ``GAOPT" implemented in BLOCK2 software.~\cite{block2}
In both cases, DMRG variational $J$ is approximately $-75$ $cm^{-1}$, while the extrapolated $J$ values differ in magnitude.
DMRG-Fiedler extrapolated $J$ is $-$105.64 cm$^{-1}$, while for DMRG-GAOPT extrapolated $J$ is $-$90.87 cm$^{-1}$.
TPSCI and DMRG energy extrapolation plots are shown in Fig. \ref{fig:Mn_complex}. The DMRG-Fiedler extrapolation curve for $S^2=0$ is notably steep as the DMRG variational energy is far from extrapolated energy because of large discarded weights. This is an indication we might need a significantly larger MPS bond-dimension ($\chi \gg 1000$) calculation to extrapolate efficiently $\chi=\infty$. For comparison, we have also computed the extrapolated energy plots for GAOPT orbital ordering. In contrast to the Fiedler ordering, the GAOPT-ordered results display a consistent convergence pattern for the high and low-spin states.
Despite the more consistent convergence, the $\Delta S_b$ value for the GAOPT-ordered results was of 21.88 cm$^{-1}$ indicating a less reliable extrapolation overall.
Standard errors in $J$ for both DMRG approaches are larger than TPSCI extrapolation errors.
Nonetheless, both DMRG approaches yield lower variational numbers than our current TPSCI calculations.
However, neither these DMRG nor TPSCI results are converged tightly enough for typical application purposes.
While the DMRG results could be readily improved by making use of multi-node parallelism, TPSCI does not yet have a distributed memory parallel implementation.

\subsection{Complex \textbf{F: }Ni-cubane}

Tetranuclear nickel-oxo assemblies are often recognized as effective catalysts for selective organic transformations, including the bromination of phenols, and exhibit redox activity relevant to energy conversion processes.~\cite{D1RA03071J,doi:10.1021/ic00120a022}
Here, we study a simplified model which contains the key features of four nickel (II) centers, 
 each with two unpaired electrons.

\subsubsection{TPSCI Energy Convergence}
 \begin{figure}[h]
    \centering
    \includegraphics[width=0.98\linewidth]{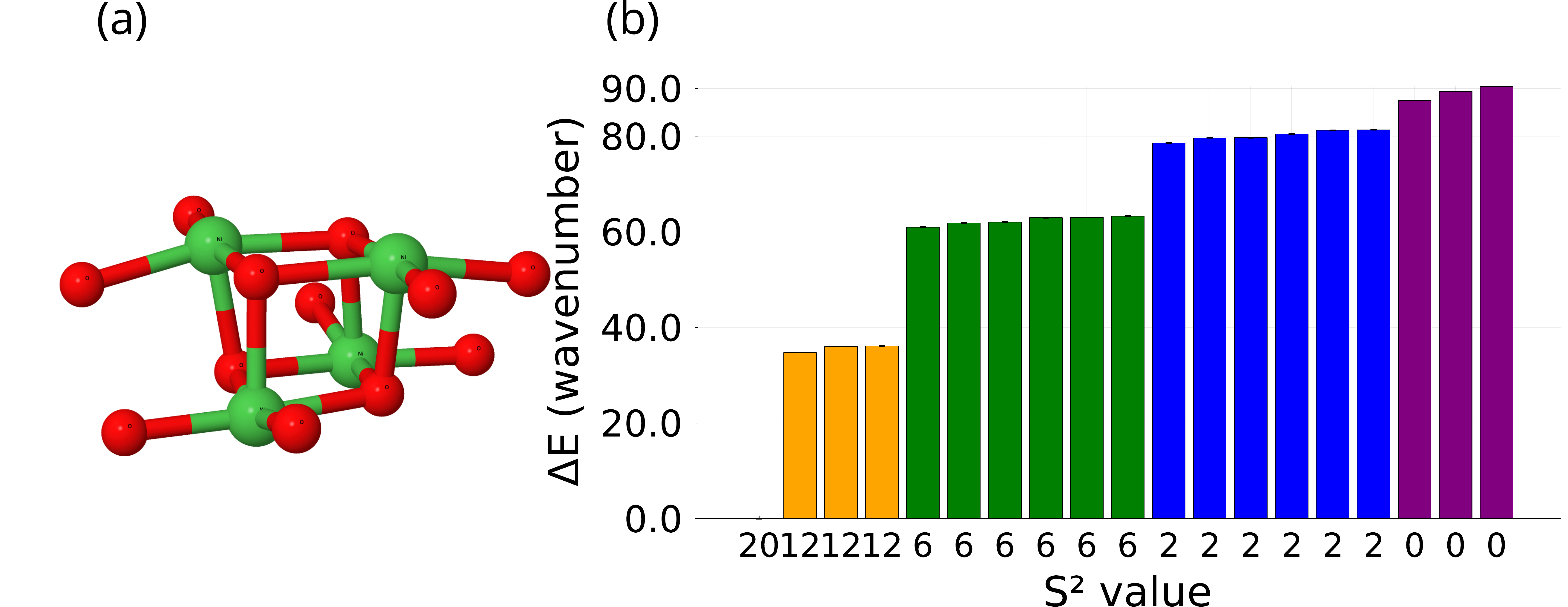}
    \caption{(a) Nickel-cubane core.
(b) Low-energy spectrum computed from the TPSCI extrapolated energies of ($S_{max}-$1) calculation + Heisenberg Diagonalization. The y-axis is the $\Delta E$ from the ground state in wavenumber. The x-axis represents $\expval{S^2}$ values for 19 low-lying spin states.}
    \label{fig:Cubane_energy}
\end{figure}
As the only ferromagnetically coupled system in this paper, the ground state for the Ni-cubane core is a high spin state with $\expval{S^2}$ = 20. 
The 8 unpaired electrons among four metal centers give rise to 19 low-lying electronic states, with $\expval{S^2}$ ranging from 0 to 20.
In contrast to the previous examples, we were not able to tightly converge the full low-energy Ni-cubane TPSCI spectrum  on a single compute node. 
Instead, we used the local character of the TPSCI basis to sidestep the direct computation, indirectly computing each of the pair-wise exchange coupling constants from the off-diagonal matrix elements of a Bloch effective Hamiltonian. 
This approach, only required that we converge 4 TPSCI eigenstates in the $M_s=3$ manifold of TPS basis vectors. 
The explicit procedure is described in Appendix \ref{j_multimetal}.
After forming the Bloch effective Hamiltonian, $H_{eff}$, we have used equation \ref{j_multicenter} to compute $J$ values between different nickel centers. 
\begin{table}[h]
    \centering
    \begin{tabular}{c|cccccc}\hline \hline
       TPSCI  &$J_{12}$  &$J_{13}$&$J_{14}$&$J_{23}$  &$J_{24}$&$J_{34}$ \\ \hline
      Variational  &4.78&4.47&4.54&4.53&4.69&4.73\\
      Extrapolated &4.63&4.24&4.41&4.36&4.48&4.61 \\\hline \hline
    \end{tabular}
    \caption{$J$ values (in wavenumber) between different nickel centers in complex \textbf{F}. $J_{12}$ refers to exchange coupling constant between two specific metal centers. }
    \label{tbl:j_ni-cubane}
\end{table}
Diagonalizing the resulting Heisenberg Hamiltonian gives us approximate descriptions of the full low-energy spectrum, 
and the relative energies of the 18 spin states ($S^2=\{12,6,2,0\}$) w.r.t the ground state are shown in Fig. \ref{fig:Cubane_energy}.


Due to the state-specific nature of the DMRG calculations, a state-by-state comparison would have required us to compute separate calculations for each spin-block, where each calculation targets multiple roots, complicating reliable convergence. 
As such, for comparing to DMRG, we focus only on the energy gap between the high-spin $S=4$ ground state and the lowest-energy $S=0$ state, $\Delta E_{(20,0)}$. 
The $\Delta E_{(20,0)}$ for TPSCI (M=100) variational and extrapolated energies are 91.12 and 87.43 $cm^{-1}$, respectively.
However, DMRG predicts  a slightly lower $\Delta E_{(20,0)}$: 81.24 $cm^{-1}$ for variational energy spectrum at $\chi$=1000 and 82.82 $cm^{-1}$ for extrapolated energy gap.
Although this agreement is relatively accurate, it is not possible to determine the source of the remaining $\approx 4$ $cm^{-1}$ difference. 
Errors from both cluster state truncation ($M=100$) and non-Heisenberg behavior could contribute to this difference, and resolving the difference is not possible without further convergence studies. 
However, based on observations from the previous examples, we expect that the primary difference is due to TPSCI under-estimating the spin re-coupling strength that stabilizes low-spin states, an assumption that could be tested with more accurate TPSCI calculations with a greater number of cluster states or simply improved cluster state definitions.

\section{Conclusion}
In this work, we have presented a systematic evaluation of TPSCI compared with DMRG calculations for a diverse set of spin-coupled transition metal complexes.
While both methods converge to exact diagonalization if enough computational resources are provided, we limit all calculations to a single compute node to provide more meaningful performance comparisons. 

We find that while DMRG consistently yields lower variational energies, the relative spin-state energies (exchange coupling constants) are generally in good agreement, indicating that much of the correlation energy that TPSCI misses is consistent among all spin states.
We also find that the multistate nature of TPSCI makes it relatively straightforward to obtain precise extrapolations for relative quantities, like exchange coupling constants. 

This work exposes a key shortcoming in our current strategy for setting up TPSCI calculations, namely the manner in which we truncate the local cluster Hilbert spaces. 
Currently, we use the $M$ lowest energy eigenstates of our RO-cMF Hamiltonian~\cite{bachhar_rocmf} to form a basis for the local Hilbert spaces.
Because the RO-cMF Hamiltonian contains no information about inter-cluster entanglement, this is a far from optimal choice. 
We have previously~\cite{abraham_selected_2020} explored using a more sophisticated approach for defining the local cluster basis based on a Schmidt-decomposition of an embedded cluster wavefunction. 
This approach does indeed provide significant improvements due to the entanglement-aware truncation. 
However, we were not able to apply this Embedded Schmidt Truncation (EST) in this paper due to the violation of local $S^2$ symmetry. 
In follow-up work, we will focus on developing improved cluster state truncation schemes that will be both spin-preserving and entanglement-aware, as well as distributed memory implementations and spin-preserving HOSVD rotations of the cluster state basis for further increasing the compactness of the variational subspaces. 

\section{Acknowledgments} 
The authors are grateful for the generous support from the  National Science Foundation (Award No. 1752612).

\section{Supporting Information}
An Excel spreadsheet containing the raw TPSCI and DMRG energies, as well as the computed magnetic coupling constants, is provided. In addition, a .txt file with the Cartesian coordinates of the TMCs studied in this work is included.

\bibliographystyle{achemso}
\bibliography{paper_library.bib}

\providecommand{\latin}[1]{#1}
\makeatletter
\providecommand{\doi}
  {\begingroup\let\do\@makeother\dospecials
  \catcode`\{=1 \catcode`\}=2 \doi@aux}
\providecommand{\doi@aux}[1]{\endgroup\texttt{#1}}
\makeatother
\providecommand*\mcitethebibliography{\thebibliography}
\csname @ifundefined\endcsname{endmcitethebibliography}
  {\let\endmcitethebibliography\endthebibliography}{}
\begin{mcitethebibliography}{82}
\providecommand*\natexlab[1]{#1}
\providecommand*\mciteSetBstSublistMode[1]{}
\providecommand*\mciteSetBstMaxWidthForm[2]{}
\providecommand*\mciteBstWouldAddEndPuncttrue
  {\def\EndOfBibitem{\unskip.}}
\providecommand*\mciteBstWouldAddEndPunctfalse
  {\let\EndOfBibitem\relax}
\providecommand*\mciteSetBstMidEndSepPunct[3]{}
\providecommand*\mciteSetBstSublistLabelBeginEnd[3]{}
\providecommand*\EndOfBibitem{}
\mciteSetBstSublistMode{f}
\mciteSetBstMaxWidthForm{subitem}{(\alph{mcitesubitemcount})}
\mciteSetBstSublistLabelBeginEnd
  {\mcitemaxwidthsubitemform\space}
  {\relax}
  {\relax}

\bibitem[Paul \latin{et~al.}(2017)Paul, Neese, and Pantazis]{oec}
Paul,~S.; Neese,~F.; Pantazis,~D.~A. Structural models of the biological
  oxygen-evolving complex: achievements{,} insights{,} and challenges for
  biomimicry. \emph{Green Chem.} \textbf{2017}, \emph{19}, 2309--2325\relax
\mciteBstWouldAddEndPuncttrue
\mciteSetBstMidEndSepPunct{\mcitedefaultmidpunct}
{\mcitedefaultendpunct}{\mcitedefaultseppunct}\relax
\EndOfBibitem
\bibitem[Einsle and Rees(2020)Einsle, and Rees]{nitrogenase}
Einsle,~O.; Rees,~D.~C. Structural Enzymology of Nitrogenase Enzymes.
  \emph{Chemical Reviews} \textbf{2020}, \emph{120}, 4969--5004, PMID:
  32538623\relax
\mciteBstWouldAddEndPuncttrue
\mciteSetBstMidEndSepPunct{\mcitedefaultmidpunct}
{\mcitedefaultendpunct}{\mcitedefaultseppunct}\relax
\EndOfBibitem
\bibitem[Wu \latin{et~al.}(2022)Wu, Liang, Zhang, Louis, and
  Wang]{energy_bimetallic}
Wu,~Q.; Liang,~S.; Zhang,~T.; Louis,~B.; Wang,~Q. Current advances in
  bimetallic catalysts for carbon dioxide hydrogenation to methanol.
  \emph{Fuel} \textbf{2022}, \emph{313}, 122963\relax
\mciteBstWouldAddEndPuncttrue
\mciteSetBstMidEndSepPunct{\mcitedefaultmidpunct}
{\mcitedefaultendpunct}{\mcitedefaultseppunct}\relax
\EndOfBibitem
\bibitem[Chilton(2022)]{molecular_magnetism}
Chilton,~N.~F. Molecular Magnetism. \emph{Annual Review of Materials Research}
  \textbf{2022}, \emph{52}, 79--101\relax
\mciteBstWouldAddEndPuncttrue
\mciteSetBstMidEndSepPunct{\mcitedefaultmidpunct}
{\mcitedefaultendpunct}{\mcitedefaultseppunct}\relax
\EndOfBibitem
\bibitem[Gil-Sepulcre and Llobet(2022)Gil-Sepulcre, and
  Llobet]{Gil-Sepulcre2022}
Gil-Sepulcre,~M.; Llobet,~A. Molecular water oxidation catalysts based on
  first-row transition metal complexes. \emph{Nature Catalysis} \textbf{2022},
  \emph{5}, 79--82\relax
\mciteBstWouldAddEndPuncttrue
\mciteSetBstMidEndSepPunct{\mcitedefaultmidpunct}
{\mcitedefaultendpunct}{\mcitedefaultseppunct}\relax
\EndOfBibitem
\bibitem[Tanabe and Nishibayashi(2022)Tanabe, and
  Nishibayashi]{TANABE2022214783}
Tanabe,~Y.; Nishibayashi,~Y. Recent advances in catalytic nitrogen fixation
  using transition metal–dinitrogen complexes under mild reaction conditions.
  \emph{Coordination Chemistry Reviews} \textbf{2022}, \emph{472}, 214783\relax
\mciteBstWouldAddEndPuncttrue
\mciteSetBstMidEndSepPunct{\mcitedefaultmidpunct}
{\mcitedefaultendpunct}{\mcitedefaultseppunct}\relax
\EndOfBibitem
\bibitem[{Szab{\'o}, Attila} and {Ostlund, Neil S.}(){Szab{\'o}, Attila}, and
  {Ostlund, Neil S.}]{szabo__attila_modern_nodate}
{Szab{\'o}, Attila}; {Ostlund, Neil S.} \emph{Modern quantum chemistry :
  introduction to advanced electronic structure theory}; Courier Corporation,
  2012\relax
\mciteBstWouldAddEndPuncttrue
\mciteSetBstMidEndSepPunct{\mcitedefaultmidpunct}
{\mcitedefaultendpunct}{\mcitedefaultseppunct}\relax
\EndOfBibitem
\bibitem[Helgaker \latin{et~al.}(2000)Helgaker, Jørgensen, and
  Olsen]{helgaker_molecular_2000-1}
Helgaker,~T.; Jørgensen,~P.; Olsen,~J. \emph{Molecular
  {Electronic}‐{Structure} {Theory}}, 1st ed.; Wiley, 2000\relax
\mciteBstWouldAddEndPuncttrue
\mciteSetBstMidEndSepPunct{\mcitedefaultmidpunct}
{\mcitedefaultendpunct}{\mcitedefaultseppunct}\relax
\EndOfBibitem
\bibitem[Hartree(1928)]{hartree_wave_1928}
Hartree,~D.~R. The {Wave} {Mechanics} of an {Atom} with a {Non}-{Coulomb}
  {Central} {Field}. {Part} {I}. {Theory} and {Methods}. \emph{Math. Proc.
  Camb. Phil. Soc.} \textbf{1928}, \emph{24}, 89--110\relax
\mciteBstWouldAddEndPuncttrue
\mciteSetBstMidEndSepPunct{\mcitedefaultmidpunct}
{\mcitedefaultendpunct}{\mcitedefaultseppunct}\relax
\EndOfBibitem
\bibitem[Shavitt and Bartlett(2009)Shavitt, and
  Bartlett]{shavitt_many-body_2009}
Shavitt,~I.; Bartlett,~R.~J. \emph{Many-{Body} {Methods} in {Chemistry} and
  {Physics}: {MBPT} and {Coupled}-{Cluster} {Theory}}, 1st ed.; Cambridge
  University Press, 2009\relax
\mciteBstWouldAddEndPuncttrue
\mciteSetBstMidEndSepPunct{\mcitedefaultmidpunct}
{\mcitedefaultendpunct}{\mcitedefaultseppunct}\relax
\EndOfBibitem
\bibitem[Korth(2017)]{korth_density_2017}
Korth,~M. Density {Functional} {Theory}: {Not} {Quite} the {Right} {Answer} for
  the {Right} {Reason} {Yet}. \emph{Angew Chem Int Ed} \textbf{2017},
  \emph{56}, 5396--5398\relax
\mciteBstWouldAddEndPuncttrue
\mciteSetBstMidEndSepPunct{\mcitedefaultmidpunct}
{\mcitedefaultendpunct}{\mcitedefaultseppunct}\relax
\EndOfBibitem
\bibitem[Cramer(2002)]{cramer2002essentials}
Cramer,~C.~J. \emph{Essentials of Computational Chemistry: Theories and
  Models}, 1st ed.; John Wiley \& Sons: West Sussex, England ; New York, 2002;
  pp xvii, 542\relax
\mciteBstWouldAddEndPuncttrue
\mciteSetBstMidEndSepPunct{\mcitedefaultmidpunct}
{\mcitedefaultendpunct}{\mcitedefaultseppunct}\relax
\EndOfBibitem
\bibitem[Li(2004)]{li_block-correlated_2004}
Li,~S. Block-correlated coupled cluster theory: {The} general formulation and
  its application to the antiferromagnetic {Heisenberg} model. \emph{The
  Journal of Chemical Physics} \textbf{2004}, \emph{120}, 5017--5026\relax
\mciteBstWouldAddEndPuncttrue
\mciteSetBstMidEndSepPunct{\mcitedefaultmidpunct}
{\mcitedefaultendpunct}{\mcitedefaultseppunct}\relax
\EndOfBibitem
\bibitem[Ma \latin{et~al.}(2011)Ma, Li~Manni, and Gagliardi]{gasscf}
Ma,~D.; Li~Manni,~G.; Gagliardi,~L. The generalized active space concept in
  multiconfigurational self-consistent field methods. \emph{The Journal of
  Chemical Physics} \textbf{2011}, \emph{135}, 044128\relax
\mciteBstWouldAddEndPuncttrue
\mciteSetBstMidEndSepPunct{\mcitedefaultmidpunct}
{\mcitedefaultendpunct}{\mcitedefaultseppunct}\relax
\EndOfBibitem
\bibitem[Jim\'enez-Hoyos and Scuseria(2015)Jim\'enez-Hoyos, and
  Scuseria]{cmf_first}
Jim\'enez-Hoyos,~C.~A.; Scuseria,~G.~E. Cluster-based mean-field and
  perturbative description of strongly correlated fermion systems: Application
  to the one- and two-dimensional Hubbard model. \emph{Phys. Rev. B}
  \textbf{2015}, \emph{92}, 085101\relax
\mciteBstWouldAddEndPuncttrue
\mciteSetBstMidEndSepPunct{\mcitedefaultmidpunct}
{\mcitedefaultendpunct}{\mcitedefaultseppunct}\relax
\EndOfBibitem
\bibitem[Parker \latin{et~al.}(2013)Parker, Seideman, Ratner, and
  Shiozaki]{parker_communication_2013}
Parker,~S.~M.; Seideman,~T.; Ratner,~M.~A.; Shiozaki,~T. Communication:
  {Active}-space decomposition for molecular dimers. \emph{The Journal of
  Chemical Physics} \textbf{2013}, \emph{139}, 021108\relax
\mciteBstWouldAddEndPuncttrue
\mciteSetBstMidEndSepPunct{\mcitedefaultmidpunct}
{\mcitedefaultendpunct}{\mcitedefaultseppunct}\relax
\EndOfBibitem
\bibitem[Parker and Shiozaki(2014)Parker, and
  Shiozaki]{parkerCommunicationActiveSpace2014}
Parker,~S.~M.; Shiozaki,~T. Communication: {Active} space decomposition with
  multiple sites: density matrix renormalization group algorithm. \emph{The
  Journal of chemical physics} \textbf{2014}, \emph{141}, 211102--211102\relax
\mciteBstWouldAddEndPuncttrue
\mciteSetBstMidEndSepPunct{\mcitedefaultmidpunct}
{\mcitedefaultendpunct}{\mcitedefaultseppunct}\relax
\EndOfBibitem
\bibitem[Hermes \latin{et~al.}(2020)Hermes, Pandharkar, and
  Gagliardi]{vlasscf_hermes}
Hermes,~M.~R.; Pandharkar,~R.; Gagliardi,~L. Variational Localized Active Space
  Self-Consistent Field Method. \emph{Journal of Chemical Theory and
  Computation} \textbf{2020}, \emph{16}, 4923--4937, PMID: 32491849\relax
\mciteBstWouldAddEndPuncttrue
\mciteSetBstMidEndSepPunct{\mcitedefaultmidpunct}
{\mcitedefaultendpunct}{\mcitedefaultseppunct}\relax
\EndOfBibitem
\bibitem[Hermes and Gagliardi(2019)Hermes, and Gagliardi]{vlasscf_dmet}
Hermes,~M.~R.; Gagliardi,~L. Multiconfigurational Self-Consistent Field Theory
  with Density Matrix Embedding: The Localized Active Space Self-Consistent
  Field Method. \emph{Journal of Chemical Theory and Computation}
  \textbf{2019}, \emph{15}, 972--986, PMID: 30620876\relax
\mciteBstWouldAddEndPuncttrue
\mciteSetBstMidEndSepPunct{\mcitedefaultmidpunct}
{\mcitedefaultendpunct}{\mcitedefaultseppunct}\relax
\EndOfBibitem
\bibitem[Parker \latin{et~al.}(2014)Parker, Seideman, Ratner, and
  Shiozaki]{parkerModelHamiltonianAnalysi2014}
Parker,~S.~M.; Seideman,~T.; Ratner,~M.~A.; Shiozaki,~T. Model Hamiltonian
  Analysis of Singlet Fission from First Principles. \emph{The Journal of
  Physical Chemistry C} \textbf{2014}, \emph{118}, 12700--12705\relax
\mciteBstWouldAddEndPuncttrue
\mciteSetBstMidEndSepPunct{\mcitedefaultmidpunct}
{\mcitedefaultendpunct}{\mcitedefaultseppunct}\relax
\EndOfBibitem
\bibitem[Pantazis(2019)]{pantazisMeetingChallengeMagnetic2019}
Pantazis,~D.~A. Meeting the {Challenge} of {Magnetic} {Coupling} in a
  {Triply}-{Bridged} {Chromium} {Dimer}: {Complementary} {Broken}-{Symmetry}
  {Density} {Functional} {Theory} and {Multireference} {Density} {Matrix}
  {Renormalization} {Group} {Perspectives}. \emph{Journal of Chemical Theory
  and Computation} \textbf{2019}, \emph{15}, 938--948, Publisher: American
  Chemical Society\relax
\mciteBstWouldAddEndPuncttrue
\mciteSetBstMidEndSepPunct{\mcitedefaultmidpunct}
{\mcitedefaultendpunct}{\mcitedefaultseppunct}\relax
\EndOfBibitem
\bibitem[Szalay \latin{et~al.}(2015)Szalay, Pfeffer, Murg, Barcza, Verstraete,
  Schneider, and Legeza]{szalayTensorProductMethod2015}
Szalay,~S.; Pfeffer,~M.; Murg,~V.; Barcza,~G.; Verstraete,~F.; Schneider,~R.;
  Legeza,~{\"O}. Tensor product methods and entanglement optimization for ab
  initio quantum chemistry. \emph{International Journal of Quantum Chemistry}
  \textbf{2015}, \emph{115}, 1342--1391\relax
\mciteBstWouldAddEndPuncttrue
\mciteSetBstMidEndSepPunct{\mcitedefaultmidpunct}
{\mcitedefaultendpunct}{\mcitedefaultseppunct}\relax
\EndOfBibitem
\bibitem[Baiardi and Reiher(2020)Baiardi, and
  Reiher]{baiardiDensityMatrixRenormalization2020}
Baiardi,~A.; Reiher,~M. The density matrix renormalization group in chemistry
  and molecular physics: {Recent} developments and new challenges. \emph{The
  Journal of Chemical Physics} \textbf{2020}, \emph{152}, 040903, Publisher:
  American Institute of Physics\relax
\mciteBstWouldAddEndPuncttrue
\mciteSetBstMidEndSepPunct{\mcitedefaultmidpunct}
{\mcitedefaultendpunct}{\mcitedefaultseppunct}\relax
\EndOfBibitem
\bibitem[Schollwöck(2011)]{schollwockDensitymatrixRenormalizationGroup2011}
Schollwöck,~U. The density-matrix renormalization group in the age of matrix
  product states. \emph{Annals of Physics} \textbf{2011}, \emph{326},
  96--192\relax
\mciteBstWouldAddEndPuncttrue
\mciteSetBstMidEndSepPunct{\mcitedefaultmidpunct}
{\mcitedefaultendpunct}{\mcitedefaultseppunct}\relax
\EndOfBibitem
\bibitem[Wouters and Van~Neck(2014)Wouters, and
  Van~Neck]{woutersDensityMatrixRenormalization2014}
Wouters,~S.; Van~Neck,~D. The density matrix renormalization group for ab
  initio quantum chemistry. \emph{The European Physical Journal D}
  \textbf{2014}, \emph{68}, 272--272\relax
\mciteBstWouldAddEndPuncttrue
\mciteSetBstMidEndSepPunct{\mcitedefaultmidpunct}
{\mcitedefaultendpunct}{\mcitedefaultseppunct}\relax
\EndOfBibitem
\bibitem[Eriksen \latin{et~al.}(2020)Eriksen, Anderson, Deustua, Ghanem, Hait,
  Hoffmann, Lee, Levine, Magoulas, Shen, Tubman, Whaley, Xu, Yao, Zhang, Alavi,
  Chan, Head-Gordon, Liu, Piecuch, Sharma, Ten-no, Umrigar, and
  Gauss]{eriksenGroundStateElectronic2020}
Eriksen,~J.~J. \latin{et~al.}  The {Ground} {State} {Electronic} {Energy} of
  {Benzene}. \emph{The Journal of Physical Chemistry Letters} \textbf{2020},
  \emph{11}, 8922--8929, Publisher: American Chemical Society\relax
\mciteBstWouldAddEndPuncttrue
\mciteSetBstMidEndSepPunct{\mcitedefaultmidpunct}
{\mcitedefaultendpunct}{\mcitedefaultseppunct}\relax
\EndOfBibitem
\bibitem[Olivares-Amaya \latin{et~al.}(2015)Olivares-Amaya, Hu, Nakatani,
  Sharma, Yang, and Chan]{olivares-amayaAbinitioDensityMatrix2015a}
Olivares-Amaya,~R.; Hu,~W.; Nakatani,~N.; Sharma,~S.; Yang,~J.; Chan,~G. K.-L.
  The ab-initio density matrix renormalization group in practice. \emph{The
  Journal of Chemical Physics} \textbf{2015}, \emph{142}, 034102, Publisher:
  American Institute of Physics\relax
\mciteBstWouldAddEndPuncttrue
\mciteSetBstMidEndSepPunct{\mcitedefaultmidpunct}
{\mcitedefaultendpunct}{\mcitedefaultseppunct}\relax
\EndOfBibitem
\bibitem[Sharma \latin{et~al.}(2014)Sharma, Sivalingam, Neese, and
  Chan]{sharmaLowenergySpectrumIronsulfur2014}
Sharma,~S.; Sivalingam,~K.; Neese,~F.; Chan,~G. K.-L. Low-energy spectrum of
  iron-sulfur clusters directly from many-particle quantum mechanics.
  \emph{Nature chemistry} \textbf{2014}, \emph{6}, 927--33\relax
\mciteBstWouldAddEndPuncttrue
\mciteSetBstMidEndSepPunct{\mcitedefaultmidpunct}
{\mcitedefaultendpunct}{\mcitedefaultseppunct}\relax
\EndOfBibitem
\bibitem[Sharma \latin{et~al.}(2016)Sharma, Jeanmairet, and
  Alavi]{sharmaQuasidegeneratePerturbationTheory2016a}
Sharma,~S.; Jeanmairet,~G.; Alavi,~A. Quasi-degenerate perturbation theory
  using matrix product states. \emph{The Journal of Chemical Physics}
  \textbf{2016}, \emph{144}, 034103\relax
\mciteBstWouldAddEndPuncttrue
\mciteSetBstMidEndSepPunct{\mcitedefaultmidpunct}
{\mcitedefaultendpunct}{\mcitedefaultseppunct}\relax
\EndOfBibitem
\bibitem[Zhai and Chan(2021)Zhai, and Chan]{zhai2021low}
Zhai,~H.; Chan,~G.~K. Low communication high performance ab initio density
  matrix renormalization group algorithms. \emph{The Journal of Chemical
  Physics} \textbf{2021}, \emph{154}\relax
\mciteBstWouldAddEndPuncttrue
\mciteSetBstMidEndSepPunct{\mcitedefaultmidpunct}
{\mcitedefaultendpunct}{\mcitedefaultseppunct}\relax
\EndOfBibitem
\bibitem[Abraham and Mayhall(2020)Abraham, and Mayhall]{abraham_selected_2020}
Abraham,~V.; Mayhall,~N.~J. Selected {Configuration} {Interaction} in a {Basis}
  of {Cluster} {State} {Tensor} {Products}. \emph{J. Chem. Theory Comput.}
  \textbf{2020}, \emph{16}, 6098--6113\relax
\mciteBstWouldAddEndPuncttrue
\mciteSetBstMidEndSepPunct{\mcitedefaultmidpunct}
{\mcitedefaultendpunct}{\mcitedefaultseppunct}\relax
\EndOfBibitem
\bibitem[Braunscheidel \latin{et~al.}(2023)Braunscheidel, Abraham, and
  Mayhall]{braunscheidel_generalization_2023}
Braunscheidel,~N.~M.; Abraham,~V.; Mayhall,~N.~J. Generalization of the
  {Tensor} {Product} {Selected} {CI} {Method} for {Molecular} {Excited}
  {States}. \emph{J. Phys. Chem. A} \textbf{2023}, \emph{127}, 8179--8193\relax
\mciteBstWouldAddEndPuncttrue
\mciteSetBstMidEndSepPunct{\mcitedefaultmidpunct}
{\mcitedefaultendpunct}{\mcitedefaultseppunct}\relax
\EndOfBibitem
\bibitem[Braunscheidel \latin{et~al.}(2024)Braunscheidel, Bachhar, and
  Mayhall]{faraday_discussions}
Braunscheidel,~N.~M.; Bachhar,~A.; Mayhall,~N.~J. Accurate and Interpretable
  Representation of Correlated Electronic Structure via Tensor Product Selected
  CI. \emph{Faraday Discuss.} \textbf{2024}, --\relax
\mciteBstWouldAddEndPuncttrue
\mciteSetBstMidEndSepPunct{\mcitedefaultmidpunct}
{\mcitedefaultendpunct}{\mcitedefaultseppunct}\relax
\EndOfBibitem
\bibitem[Papastathopoulos-Katsaros
  \latin{et~al.}(2023)Papastathopoulos-Katsaros, Henderson, and
  Scuseria]{cmf_third}
Papastathopoulos-Katsaros,~A.; Henderson,~T.~M.; Scuseria,~G.~E.
  {Symmetry-projected cluster mean-field theory applied to spin systems}.
  \emph{The Journal of Chemical Physics} \textbf{2023}, \emph{159},
  084107\relax
\mciteBstWouldAddEndPuncttrue
\mciteSetBstMidEndSepPunct{\mcitedefaultmidpunct}
{\mcitedefaultendpunct}{\mcitedefaultseppunct}\relax
\EndOfBibitem
\bibitem[Papastathopoulos-Katsaros
  \latin{et~al.}(2024)Papastathopoulos-Katsaros, Henderson, and
  Scuseria]{cMF_linear_combinations}
Papastathopoulos-Katsaros,~A.; Henderson,~T.~M.; Scuseria,~G.~E. Linear
  Combinations of Cluster Mean-Field States Applied to Spin Systems.
  \emph{Journal of Chemical Theory and Computation} \textbf{2024}, \emph{20},
  3697--3705, PMID: 38695526\relax
\mciteBstWouldAddEndPuncttrue
\mciteSetBstMidEndSepPunct{\mcitedefaultmidpunct}
{\mcitedefaultendpunct}{\mcitedefaultseppunct}\relax
\EndOfBibitem
\bibitem[Bachhar and Mayhall(2024)Bachhar, and Mayhall]{bachhar_rocmf}
Bachhar,~A.; Mayhall,~N.~J. Restricted Open-Shell Cluster Mean-Field theory for
  Strongly Correlated Systems. \emph{The Journal of Physical Chemistry A}
  \textbf{2024}, \emph{128}, 9015--9027, PMID: 39373627\relax
\mciteBstWouldAddEndPuncttrue
\mciteSetBstMidEndSepPunct{\mcitedefaultmidpunct}
{\mcitedefaultendpunct}{\mcitedefaultseppunct}\relax
\EndOfBibitem
\bibitem[Huron \latin{et~al.}(1973)Huron, Malrieu, and
  Rancurel]{huron1973iterative}
Huron,~B.; Malrieu,~J.; Rancurel,~P. Iterative perturbation calculations of
  ground and excited state energies from multiconfigurational zeroth-order
  wavefunctions. \emph{The Journal of Chemical Physics} \textbf{1973},
  \emph{58}, 5745--5759\relax
\mciteBstWouldAddEndPuncttrue
\mciteSetBstMidEndSepPunct{\mcitedefaultmidpunct}
{\mcitedefaultendpunct}{\mcitedefaultseppunct}\relax
\EndOfBibitem
\bibitem[Evangelisti \latin{et~al.}(1983)Evangelisti, Daudey, and
  Malrieu]{evangelisti1983convergence}
Evangelisti,~S.; Daudey,~J.-P.; Malrieu,~J.-P. Convergence of an improved CIPSI
  algorithm. \emph{Chemical Physics} \textbf{1983}, \emph{75}, 91--102\relax
\mciteBstWouldAddEndPuncttrue
\mciteSetBstMidEndSepPunct{\mcitedefaultmidpunct}
{\mcitedefaultendpunct}{\mcitedefaultseppunct}\relax
\EndOfBibitem
\bibitem[Loos \latin{et~al.}(2018)Loos, Scemama, Blondel, Garniron, Caffarel,
  and Jacquemin]{loos2018mountaineering}
Loos,~P.-F.; Scemama,~A.; Blondel,~A.; Garniron,~Y.; Caffarel,~M.;
  Jacquemin,~D. A mountaineering strategy to excited states: Highly accurate
  reference energies and benchmarks. \emph{Journal of chemical theory and
  computation} \textbf{2018}, \emph{14}, 4360--4379\relax
\mciteBstWouldAddEndPuncttrue
\mciteSetBstMidEndSepPunct{\mcitedefaultmidpunct}
{\mcitedefaultendpunct}{\mcitedefaultseppunct}\relax
\EndOfBibitem
\bibitem[Abraham and Mayhall(2021)Abraham, and Mayhall]{abraham_cluster_2021}
Abraham,~V.; Mayhall,~N.~J. Cluster many-body expansion: {A} many-body
  expansion of the electron correlation energy about a cluster mean field
  reference. \emph{The Journal of Chemical Physics} \textbf{2021}, \emph{155},
  054101\relax
\mciteBstWouldAddEndPuncttrue
\mciteSetBstMidEndSepPunct{\mcitedefaultmidpunct}
{\mcitedefaultendpunct}{\mcitedefaultseppunct}\relax
\EndOfBibitem
\bibitem[Abraham and Mayhall(2022)Abraham, and Mayhall]{abraham_coupled_2022}
Abraham,~V.; Mayhall,~N.~J. Coupled {Electron} {Pair}-{Type} {Approximations}
  for {Tensor} {Product} {State} {Wave} {Functions}. \emph{J. Chem. Theory
  Comput.} \textbf{2022}, \emph{18}, 4856--4864\relax
\mciteBstWouldAddEndPuncttrue
\mciteSetBstMidEndSepPunct{\mcitedefaultmidpunct}
{\mcitedefaultendpunct}{\mcitedefaultseppunct}\relax
\EndOfBibitem
\bibitem[Holmes \latin{et~al.}(2016)Holmes, Tubman, and
  Umrigar]{holmesHeatBathCI2016}
Holmes,~A.~A.; Tubman,~N.~M.; Umrigar,~C.~J. Heat-{Bath} {Configuration}
  {Interaction}: {An} {Efficient} {Selected} {Configuration} {Interaction}
  {Algorithm} {Inspired} by {Heat}-{Bath} {Sampling}. \emph{Journal of Chemical
  Theory and Computation} \textbf{2016}, \emph{12}, 3674--3680\relax
\mciteBstWouldAddEndPuncttrue
\mciteSetBstMidEndSepPunct{\mcitedefaultmidpunct}
{\mcitedefaultendpunct}{\mcitedefaultseppunct}\relax
\EndOfBibitem
\bibitem[Note1()]{Note1}
For systems away from half-filling, alternative models such as the $tJ$ or
  double exchange Hamiltonians may be more appropriate.\relax
\mciteBstWouldAddEndPunctfalse
\mciteSetBstMidEndSepPunct{\mcitedefaultmidpunct}
{}{\mcitedefaultseppunct}\relax
\EndOfBibitem
\bibitem[Malrieu \latin{et~al.}(2014)Malrieu, Caballol, Calzado, de~Graaf, and
  Guihéry]{malrieuMagneticInteractionsMolecules2014}
Malrieu,~J.~P.; Caballol,~R.; Calzado,~C.~J.; de~Graaf,~C.; Guihéry,~N.
  Magnetic {Interactions} in {Molecules} and {Highly} {Correlated} {Materials}:
  {Physical} {Content}, {Analytical} {Derivation}, and {Rigorous} {Extraction}
  of {Magnetic} {Hamiltonians}. \emph{Chemical Reviews} \textbf{2014},
  \emph{114}, 429--492\relax
\mciteBstWouldAddEndPuncttrue
\mciteSetBstMidEndSepPunct{\mcitedefaultmidpunct}
{\mcitedefaultendpunct}{\mcitedefaultseppunct}\relax
\EndOfBibitem
\bibitem[Harris \latin{et~al.}(2014)Harris, Kurashige, Yanai, and
  Morokuma]{harrisInitioDensityMatrix2014}
Harris,~T.~V.; Kurashige,~Y.; Yanai,~T.; Morokuma,~K. Ab initio density matrix
  renormalization group study of magnetic coupling in dinuclear iron and
  chromium complexes. \emph{The Journal of Chemical Physics} \textbf{2014},
  \emph{140}, 054303, Publisher: American Institute of Physics\relax
\mciteBstWouldAddEndPuncttrue
\mciteSetBstMidEndSepPunct{\mcitedefaultmidpunct}
{\mcitedefaultendpunct}{\mcitedefaultseppunct}\relax
\EndOfBibitem
\bibitem[Mayhall and Head-Gordon(2014)Mayhall, and
  Head-Gordon]{mayhallComputationalQuantumChemistry2014}
Mayhall,~N.~J.; Head-Gordon,~M. Computational quantum chemistry for single
  {Heisenberg} spin couplings made simple: {Just} one spin flip required.
  \emph{The Journal of Chemical Physics} \textbf{2014}, \emph{141},
  134111\relax
\mciteBstWouldAddEndPuncttrue
\mciteSetBstMidEndSepPunct{\mcitedefaultmidpunct}
{\mcitedefaultendpunct}{\mcitedefaultseppunct}\relax
\EndOfBibitem
\bibitem[Mayhall and Head-Gordon(2015)Mayhall, and
  Head-Gordon]{mayhallComputationalQuantumChemistry2015a}
Mayhall,~N.~J.; Head-Gordon,~M. Computational {Quantum} {Chemistry} for
  {Multiple}-{Site} {Heisenberg} {Spin} {Couplings} {Made} {Simple}: {Still}
  {Only} {One} {Spin}–{Flip} {Required}. \emph{The Journal of Physical
  Chemistry Letters} \textbf{2015}, \emph{6}, 1982--1988\relax
\mciteBstWouldAddEndPuncttrue
\mciteSetBstMidEndSepPunct{\mcitedefaultmidpunct}
{\mcitedefaultendpunct}{\mcitedefaultseppunct}\relax
\EndOfBibitem
\bibitem[Mayhall \latin{et~al.}(2014)Mayhall, Horn, Sundstrom, and
  Head-Gordon]{mayhallSpinflipNonorthogonalConfiguration2014a}
Mayhall,~N.~J.; Horn,~P.~R.; Sundstrom,~E.~J.; Head-Gordon,~M. Spin-flip
  non-orthogonal configuration interaction: a variational and almost black-box
  method for describing strongly correlated molecules. \emph{Physical chemistry
  chemical physics : PCCP} \textbf{2014}, \emph{16}, 22694--22705\relax
\mciteBstWouldAddEndPuncttrue
\mciteSetBstMidEndSepPunct{\mcitedefaultmidpunct}
{\mcitedefaultendpunct}{\mcitedefaultseppunct}\relax
\EndOfBibitem
\bibitem[Pokhilko and Krylov(2020)Pokhilko, and
  Krylov]{pokhilkoEffectiveHamiltoniansDerived2020}
Pokhilko,~P.; Krylov,~A.~I. Effective {Hamiltonians} derived from
  equation-of-motion coupled-cluster wave functions: {Theory} and application
  to the {Hubbard} and {Heisenberg} {Hamiltonians}. \emph{The Journal of
  Chemical Physics} \textbf{2020}, \emph{152}, 094108\relax
\mciteBstWouldAddEndPuncttrue
\mciteSetBstMidEndSepPunct{\mcitedefaultmidpunct}
{\mcitedefaultendpunct}{\mcitedefaultseppunct}\relax
\EndOfBibitem
\bibitem[Houck and Mayhall(2019)Houck, and
  Mayhall]{houckCombinedSpinFlipIP2019}
Houck,~S.~E.; Mayhall,~N.~J. A Combined Spin-Flip and IP/EA Approach for
  Handling Spin and Spatial Degeneracies: Application to Double Exchange
  Systems. \emph{Journal of Chemical Theory and Computation} \textbf{2019},
  \emph{15}, 2278--2290, PMID: 30802408\relax
\mciteBstWouldAddEndPuncttrue
\mciteSetBstMidEndSepPunct{\mcitedefaultmidpunct}
{\mcitedefaultendpunct}{\mcitedefaultseppunct}\relax
\EndOfBibitem
\bibitem[Sun \latin{et~al.}(2020)Sun, Zhang, Banerjee, Bao, Barbry, Blunt,
  Bogdanov, Booth, Chen, Cui, Eriksen, Gao, Guo, Hermann, Hermes, Koh, Koval,
  Lehtola, Li, Liu, Mardirossian, McClain, Motta, Mussard, Pham, Pulkin,
  Purwanto, Robinson, Ronca, Sayfutyarova, Scheurer, Schurkus, Smith, Sun, Sun,
  Upadhyay, Wagner, Wang, White, Whitfield, Williamson, Wouters, Yang, Yu, Zhu,
  Berkelbach, Sharma, Sokolov, and Chan]{sun_recent_2020}
Sun,~Q. \latin{et~al.}  Recent developments in the {PySCF} program package.
  \emph{The Journal of Chemical Physics} \textbf{2020}, \emph{153},
  024109\relax
\mciteBstWouldAddEndPuncttrue
\mciteSetBstMidEndSepPunct{\mcitedefaultmidpunct}
{\mcitedefaultendpunct}{\mcitedefaultseppunct}\relax
\EndOfBibitem
\bibitem[{Mayhall,N. J.; Abraham, V.; Braunscheidel, N. M.; Bachhar,
  A.}()]{mayhall_nicholas_fermicg_nodate}
{Mayhall,N. J.; Abraham, V.; Braunscheidel, N. M.; Bachhar, A.} {FermiCG, 2023,
  https://github.com/mayhallgroup/FermiCG (accessed 11-24-2023)}.
  \url{https://github.com/nmayhall-vt/FermiCG.jl}\relax
\mciteBstWouldAddEndPuncttrue
\mciteSetBstMidEndSepPunct{\mcitedefaultmidpunct}
{\mcitedefaultendpunct}{\mcitedefaultseppunct}\relax
\EndOfBibitem
\bibitem[Zhai \latin{et~al.}(2023)Zhai, Larsson, Lee, Cui, Zhu, Sun, Peng,
  Peng, Liao, Tölle, Yang, Li, and Chan]{block2}
Zhai,~H.; Larsson,~H.~R.; Lee,~S.; Cui,~Z.-H.; Zhu,~T.; Sun,~C.; Peng,~L.;
  Peng,~R.; Liao,~K.; Tölle,~J.; Yang,~J.; Li,~S.; Chan,~G. K.-L. {Block2: A
  comprehensive open source framework to develop and apply state-of-the-art
  DMRG algorithms in electronic structure and beyond}. \emph{The Journal of
  Chemical Physics} \textbf{2023}, \emph{159}, 234801\relax
\mciteBstWouldAddEndPuncttrue
\mciteSetBstMidEndSepPunct{\mcitedefaultmidpunct}
{\mcitedefaultendpunct}{\mcitedefaultseppunct}\relax
\EndOfBibitem
\bibitem[Niemann \latin{et~al.}(1992)Niemann, Bossek, Wieghardt, Butzlaff,
  Trautwein, and Nuber]{cr2_pantazis_experimental}
Niemann,~A.; Bossek,~U.; Wieghardt,~K.; Butzlaff,~C.; Trautwein,~A.~X.;
  Nuber,~B. A New Structure–Magnetism Relationship for Face-Sharing
  Transition-Metal Complexes with d3–d3 Electronic Configuration.
  \emph{Angewandte Chemie International Edition in English} \textbf{1992},
  \emph{31}, 311--313\relax
\mciteBstWouldAddEndPuncttrue
\mciteSetBstMidEndSepPunct{\mcitedefaultmidpunct}
{\mcitedefaultendpunct}{\mcitedefaultseppunct}\relax
\EndOfBibitem
\bibitem[Bender and Davidson(1969)Bender, and Davidson]{bender1969studies}
Bender,~C.~F.; Davidson,~E.~R. Studies in configuration interaction: The
  first-row diatomic hydrides. \emph{Physical Review} \textbf{1969},
  \emph{183}, 23\relax
\mciteBstWouldAddEndPuncttrue
\mciteSetBstMidEndSepPunct{\mcitedefaultmidpunct}
{\mcitedefaultendpunct}{\mcitedefaultseppunct}\relax
\EndOfBibitem
\bibitem[Whitten and Hackmeyer(1969)Whitten, and
  Hackmeyer]{whitten1969configuration}
Whitten,~J.; Hackmeyer,~M. Configuration interaction studies of ground and
  excited states of polyatomic molecules. I. The CI formulation and studies of
  formaldehyde. \emph{The Journal of Chemical Physics} \textbf{1969},
  \emph{51}, 5584--5596\relax
\mciteBstWouldAddEndPuncttrue
\mciteSetBstMidEndSepPunct{\mcitedefaultmidpunct}
{\mcitedefaultendpunct}{\mcitedefaultseppunct}\relax
\EndOfBibitem
\bibitem[Tubman \latin{et~al.}(2016)Tubman, Lee, Takeshita, Head-Gordon, and
  Whaley]{tubman2016deterministic}
Tubman,~N.~M.; Lee,~J.; Takeshita,~T.~Y.; Head-Gordon,~M.; Whaley,~K.~B. A
  deterministic alternative to the full configuration interaction quantum Monte
  Carlo method. \emph{The Journal of chemical physics} \textbf{2016},
  \emph{145}\relax
\mciteBstWouldAddEndPuncttrue
\mciteSetBstMidEndSepPunct{\mcitedefaultmidpunct}
{\mcitedefaultendpunct}{\mcitedefaultseppunct}\relax
\EndOfBibitem
\bibitem[Schriber and Evangelista(2016)Schriber, and
  Evangelista]{schriber2016communication}
Schriber,~J.~B.; Evangelista,~F.~A. Communication: An adaptive configuration
  interaction approach for strongly correlated electrons with tunable accuracy.
  \emph{The Journal of chemical physics} \textbf{2016}, \emph{144}\relax
\mciteBstWouldAddEndPuncttrue
\mciteSetBstMidEndSepPunct{\mcitedefaultmidpunct}
{\mcitedefaultendpunct}{\mcitedefaultseppunct}\relax
\EndOfBibitem
\bibitem[Holmes \latin{et~al.}(2016)Holmes, Tubman, and
  Umrigar]{holmes2016heat}
Holmes,~A.~A.; Tubman,~N.~M.; Umrigar,~C. Heat-bath configuration interaction:
  An efficient selected configuration interaction algorithm inspired by
  heat-bath sampling. \emph{Journal of chemical theory and computation}
  \textbf{2016}, \emph{12}, 3674--3680\relax
\mciteBstWouldAddEndPuncttrue
\mciteSetBstMidEndSepPunct{\mcitedefaultmidpunct}
{\mcitedefaultendpunct}{\mcitedefaultseppunct}\relax
\EndOfBibitem
\bibitem[Liu and Hoffmann(2016)Liu, and Hoffmann]{liu2016ici}
Liu,~W.; Hoffmann,~M.~R. iCI: Iterative CI toward full CI. \emph{Journal of
  chemical theory and computation} \textbf{2016}, \emph{12}, 1169--1178\relax
\mciteBstWouldAddEndPuncttrue
\mciteSetBstMidEndSepPunct{\mcitedefaultmidpunct}
{\mcitedefaultendpunct}{\mcitedefaultseppunct}\relax
\EndOfBibitem
\bibitem[Chakraborty \latin{et~al.}(2018)Chakraborty, Ghosh, and
  Ghosh]{chakraborty2018evolutionary}
Chakraborty,~R.; Ghosh,~P.; Ghosh,~D. Evolutionary algorithm based
  configuration interaction approach. \emph{International Journal of Quantum
  Chemistry} \textbf{2018}, \emph{118}, e25509\relax
\mciteBstWouldAddEndPuncttrue
\mciteSetBstMidEndSepPunct{\mcitedefaultmidpunct}
{\mcitedefaultendpunct}{\mcitedefaultseppunct}\relax
\EndOfBibitem
\bibitem[Ohtsuka and Hasegawa(2017)Ohtsuka, and Hasegawa]{ohtsuka2017selected}
Ohtsuka,~Y.; Hasegawa,~J.-y. Selected configuration interaction method using
  sampled first-order corrections to wave functions. \emph{The Journal of
  Chemical Physics} \textbf{2017}, \emph{147}\relax
\mciteBstWouldAddEndPuncttrue
\mciteSetBstMidEndSepPunct{\mcitedefaultmidpunct}
{\mcitedefaultendpunct}{\mcitedefaultseppunct}\relax
\EndOfBibitem
\bibitem[Levine \latin{et~al.}(2020)Levine, Hait, Tubman, Lehtola, Whaley, and
  Head-Gordon]{levine2020casscf}
Levine,~D.~S.; Hait,~D.; Tubman,~N.~M.; Lehtola,~S.; Whaley,~K.~B.;
  Head-Gordon,~M. CASSCF with extremely large active spaces using the adaptive
  sampling configuration interaction method. \emph{Journal of chemical theory
  and computation} \textbf{2020}, \emph{16}, 2340--2354\relax
\mciteBstWouldAddEndPuncttrue
\mciteSetBstMidEndSepPunct{\mcitedefaultmidpunct}
{\mcitedefaultendpunct}{\mcitedefaultseppunct}\relax
\EndOfBibitem
\bibitem[Kossoski and Loos(2023)Kossoski, and Loos]{kossoski2023seniority}
Kossoski,~F.; Loos,~P.-F. Seniority and hierarchy configuration interaction for
  radicals and excited states. \emph{Journal of Chemical Theory and
  Computation} \textbf{2023}, \emph{19}, 8654--8670\relax
\mciteBstWouldAddEndPuncttrue
\mciteSetBstMidEndSepPunct{\mcitedefaultmidpunct}
{\mcitedefaultendpunct}{\mcitedefaultseppunct}\relax
\EndOfBibitem
\bibitem[Kossoski and Loos(2023)Kossoski, and Loos]{kossoski2023state}
Kossoski,~F.; Loos,~P.-F. State-specific configuration interaction for excited
  states. \emph{Journal of Chemical Theory and Computation} \textbf{2023},
  \emph{19}, 2258--2269\relax
\mciteBstWouldAddEndPuncttrue
\mciteSetBstMidEndSepPunct{\mcitedefaultmidpunct}
{\mcitedefaultendpunct}{\mcitedefaultseppunct}\relax
\EndOfBibitem
\bibitem[Kossoski \latin{et~al.}(2022)Kossoski, Damour, and
  Loos]{kossoski2022hierarchy}
Kossoski,~F.; Damour,~Y.; Loos,~P.-F. Hierarchy configuration interaction:
  Combining seniority number and excitation degree. \emph{The journal of
  physical chemistry letters} \textbf{2022}, \emph{13}, 4342--4349\relax
\mciteBstWouldAddEndPuncttrue
\mciteSetBstMidEndSepPunct{\mcitedefaultmidpunct}
{\mcitedefaultendpunct}{\mcitedefaultseppunct}\relax
\EndOfBibitem
\bibitem[Chilkuri \latin{et~al.}(2021)Chilkuri, Applencourt, Gasperich, Loos,
  and Scemama]{chilkuri2021spin}
Chilkuri,~V.~G.; Applencourt,~T.; Gasperich,~K.; Loos,~P.-F.; Scemama,~A.
  \emph{Advances in Quantum Chemistry}; Elsevier, 2021; Vol.~83; pp
  65--81\relax
\mciteBstWouldAddEndPuncttrue
\mciteSetBstMidEndSepPunct{\mcitedefaultmidpunct}
{\mcitedefaultendpunct}{\mcitedefaultseppunct}\relax
\EndOfBibitem
\bibitem[Loos \latin{et~al.}(2020)Loos, Damour, and
  Scemama]{loos2020performance}
Loos,~P.-F.; Damour,~Y.; Scemama,~A. The performance of CIPSI on the ground
  state electronic energy of benzene. \emph{The Journal of Chemical Physics}
  \textbf{2020}, \emph{153}\relax
\mciteBstWouldAddEndPuncttrue
\mciteSetBstMidEndSepPunct{\mcitedefaultmidpunct}
{\mcitedefaultendpunct}{\mcitedefaultseppunct}\relax
\EndOfBibitem
\bibitem[Garniron \latin{et~al.}(2018)Garniron, Scemama, Giner, Caffarel, and
  Loos]{garniron2018selected}
Garniron,~Y.; Scemama,~A.; Giner,~E.; Caffarel,~M.; Loos,~P.-F. Selected
  configuration interaction dressed by perturbation. \emph{The Journal of
  Chemical Physics} \textbf{2018}, \emph{149}\relax
\mciteBstWouldAddEndPuncttrue
\mciteSetBstMidEndSepPunct{\mcitedefaultmidpunct}
{\mcitedefaultendpunct}{\mcitedefaultseppunct}\relax
\EndOfBibitem
\bibitem[Burton and Loos(2024)Burton, and Loos]{extrapolation_selected_ci}
Burton,~H. G.~A.; Loos,~P.-F. Rationale for the extrapolation procedure in
  selected configuration interaction. \emph{The Journal of Chemical Physics}
  \textbf{2024}, \emph{160}, 104102\relax
\mciteBstWouldAddEndPuncttrue
\mciteSetBstMidEndSepPunct{\mcitedefaultmidpunct}
{\mcitedefaultendpunct}{\mcitedefaultseppunct}\relax
\EndOfBibitem
\bibitem[Ren and Chen(2005)Ren, and Chen]{REN2005159}
Ren,~Q.; Chen,~Z. Comparative study on effects of bridging and terminal ligands
  on magnetic exchange interaction in
  [(NH$_3$)$_5$Cr((\textmu-X)Cr(NH$_3$)$_4$L]$^{n+}$ (X = O, OH; L = OH,
  OH$_2$, NH$_3$; $n$ = 4, 5): density functional theory study. \emph{Journal
  of Molecular Structure: THEOCHEM} \textbf{2005}, \emph{719}, 159--168\relax
\mciteBstWouldAddEndPuncttrue
\mciteSetBstMidEndSepPunct{\mcitedefaultmidpunct}
{\mcitedefaultendpunct}{\mcitedefaultseppunct}\relax
\EndOfBibitem
\bibitem[Harris \latin{et~al.}(2014)Harris, Kurashige, Yanai, and
  Morokuma]{harris_ab_2014}
Harris,~T.~V.; Kurashige,~Y.; Yanai,~T.; Morokuma,~K. \textit{{Ab} initio}
  density matrix renormalization group study of magnetic coupling in dinuclear
  iron and chromium complexes. \emph{The Journal of Chemical Physics}
  \textbf{2014}, \emph{140}, 054303\relax
\mciteBstWouldAddEndPuncttrue
\mciteSetBstMidEndSepPunct{\mcitedefaultmidpunct}
{\mcitedefaultendpunct}{\mcitedefaultseppunct}\relax
\EndOfBibitem
\bibitem[Hart \latin{et~al.}(1992)Hart, Rappe, Gorun, and Upton]{uhf_fe}
Hart,~J.~R.; Rappe,~A.~K.; Gorun,~S.~M.; Upton,~T.~H. Ab initio calculation of
  the magnetic exchange interactions in (mu-oxo)diiron(III) systems using a
  broken symmetry wave function. \emph{Inorganic Chemistry} \textbf{1992},
  \emph{31}, 5254--5259\relax
\mciteBstWouldAddEndPuncttrue
\mciteSetBstMidEndSepPunct{\mcitedefaultmidpunct}
{\mcitedefaultendpunct}{\mcitedefaultseppunct}\relax
\EndOfBibitem
\bibitem[Lledós \latin{et~al.}(2003)Lledós, Moreno-Mañas, Sodupe,
  Vallribera, Mata, Martínez, and Molins]{dft_fe1}
Lledós,~A.; Moreno-Mañas,~M.; Sodupe,~M.; Vallribera,~A.; Mata,~I.;
  Martínez,~B.; Molins,~E. Bent and Linear Forms of the
  (Oxo)bis[trichloroferrate(III)] Dianion: An Intermolecular Effect Structural,
  Electronic and Magnetic Properties. \emph{European Journal of Inorganic
  Chemistry} \textbf{2003}, \emph{2003}, 4187--4194\relax
\mciteBstWouldAddEndPuncttrue
\mciteSetBstMidEndSepPunct{\mcitedefaultmidpunct}
{\mcitedefaultendpunct}{\mcitedefaultseppunct}\relax
\EndOfBibitem
\bibitem[Chen \latin{et~al.}(2001)Chen, Xu, Zhang, Yan, and Lin]{dft_fe2}
Chen,~Z.; Xu,~Z.; Zhang,~L.; Yan,~F.; Lin,~Z. Magnetic Exchange Interactions in
  Oxo-Bridged Diiron(III) Systems: Density Functional Calculations Coupling the
  Broken Symmetry Approach. \emph{The Journal of Physical Chemistry A}
  \textbf{2001}, \emph{105}, 9710--9716\relax
\mciteBstWouldAddEndPuncttrue
\mciteSetBstMidEndSepPunct{\mcitedefaultmidpunct}
{\mcitedefaultendpunct}{\mcitedefaultseppunct}\relax
\EndOfBibitem
\bibitem[Wang \latin{et~al.}(2005)Wang, Wei, Wang, and Chen]{ic_mrci}
Wang,~B.; Wei,~H.; Wang,~M.; Chen,~Z. {Ab initio multireference
  configuration-interaction theoretical study on the low-lying spin states in
  binuclear transition-metal complex: Magnetic exchange of }. \emph{The Journal
  of Chemical Physics} \textbf{2005}, \emph{122}, 204310\relax
\mciteBstWouldAddEndPuncttrue
\mciteSetBstMidEndSepPunct{\mcitedefaultmidpunct}
{\mcitedefaultendpunct}{\mcitedefaultseppunct}\relax
\EndOfBibitem
\bibitem[Tian \latin{et~al.}(2017)Tian, Li, Zhen, Zhang, and Lu]{fe2s2_1}
Tian,~B.; Li,~Z.; Zhen,~W.; Zhang,~X.; Lu,~G. Fe2S2 nano-clusters catalyze
  water splitting by removing formed oxygen using aid of an artificial gill
  under visible light. \emph{Journal of Catalysis} \textbf{2017}, \emph{352},
  572--578\relax
\mciteBstWouldAddEndPuncttrue
\mciteSetBstMidEndSepPunct{\mcitedefaultmidpunct}
{\mcitedefaultendpunct}{\mcitedefaultseppunct}\relax
\EndOfBibitem
\bibitem[Wang \latin{et~al.}(2023)Wang, Zhang, Zhang, Liu, Shang, Su, Liang,
  Wang, Ma, Li, and Liu]{fe2s2_2}
Wang,~M.; Zhang,~Z.; Zhang,~S.; Liu,~W.; Shang,~W.; Su,~X.; Liang,~Y.;
  Wang,~F.; Ma,~X.; Li,~Y.; Liu,~Y. Non-planar Nest-like [Fe2S2] Cluster Sites
  for Efficient Oxygen Reduction Catalysis. \emph{Angewandte Chemie
  International Edition} \textbf{2023}, \emph{62}, e202300826\relax
\mciteBstWouldAddEndPuncttrue
\mciteSetBstMidEndSepPunct{\mcitedefaultmidpunct}
{\mcitedefaultendpunct}{\mcitedefaultseppunct}\relax
\EndOfBibitem
\bibitem[Ebrahimi \latin{et~al.}(2022)Ebrahimi, Ciofi-Baffoni, Hagedoorn,
  Nicolet, Brun, Hagen, and Armstrong]{fe2s2_3}
Ebrahimi,~K.~H.; Ciofi-Baffoni,~S.; Hagedoorn,~P.-L.; Nicolet,~Y.; Brun,~N.
  E.~L.; Hagen,~W.~R.; Armstrong,~F.~A. Iron--sulfur clusters as inhibitors and
  catalysts of viral replication. \emph{Nature Chemistry} \textbf{2022},
  \emph{14}, 253--266\relax
\mciteBstWouldAddEndPuncttrue
\mciteSetBstMidEndSepPunct{\mcitedefaultmidpunct}
{\mcitedefaultendpunct}{\mcitedefaultseppunct}\relax
\EndOfBibitem
\bibitem[Stein \latin{et~al.}(2019)Stein, Pantazis, and
  Krewald]{stein2019orbital}
Stein,~C.~J.; Pantazis,~D.~A.; Krewald,~V. Orbital entanglement analysis of
  exchange-coupled systems. \emph{The journal of physical chemistry letters}
  \textbf{2019}, \emph{10}, 6762--6770\relax
\mciteBstWouldAddEndPuncttrue
\mciteSetBstMidEndSepPunct{\mcitedefaultmidpunct}
{\mcitedefaultendpunct}{\mcitedefaultseppunct}\relax
\EndOfBibitem
\bibitem[Kushvaha \latin{et~al.}(2021)Kushvaha, Francis, Kumar, Nag,
  Ravichandran, Roy, and Chandra~Mondal]{D1RA03071J}
Kushvaha,~S.~K.; Francis,~M.; Kumar,~J.; Nag,~E.; Ravichandran,~P.; Roy,~S.;
  Chandra~Mondal,~K. Synthesis{,} oligomerization and catalytic studies of a
  redox-active Ni4-cubane: a detailed mechanistic investigation. \emph{RSC
  Adv.} \textbf{2021}, \emph{11}, 22849--22858\relax
\mciteBstWouldAddEndPuncttrue
\mciteSetBstMidEndSepPunct{\mcitedefaultmidpunct}
{\mcitedefaultendpunct}{\mcitedefaultseppunct}\relax
\EndOfBibitem
\bibitem[Halcrow \latin{et~al.}(1995)Halcrow, Sun, Huffman, and
  Christou]{doi:10.1021/ic00120a022}
Halcrow,~M.~A.; Sun,~J.-S.; Huffman,~J.~C.; Christou,~G. Structural and
  Magnetic Properties of [Ni4(.mu.3-OMe)4(dbm)4(MeOH)4] and
  [Ni4(.eta.1,.mu.3-N3)4(dbm)4(EtOH)4]. Magnetostructural Correlations for
  [Ni4X4]4+ Cubane Complexes. \emph{Inorganic Chemistry} \textbf{1995},
  \emph{34}, 4167--4177\relax
\mciteBstWouldAddEndPuncttrue
\mciteSetBstMidEndSepPunct{\mcitedefaultmidpunct}
{\mcitedefaultendpunct}{\mcitedefaultseppunct}\relax
\EndOfBibitem
\end{mcitethebibliography}
\appendix
\section{Molecular Orbitals of Active space and Energy extrapolation plots}
\subsection{Complex \textbf{A: }$[\text{L}_2\text{Cr(III)}_2(\mu-\text{OH})_3]^{3+}$}\label{cr2_appendix}
\begin{figure*}
    \centering
    \includegraphics[width=0.95\linewidth]{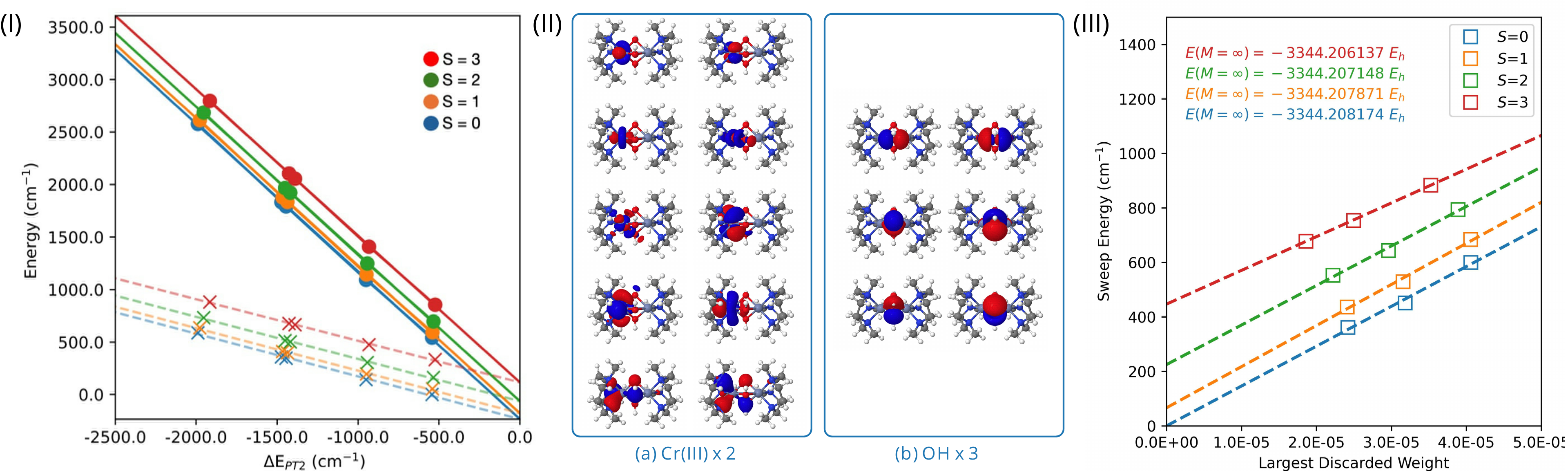}
    \caption{(I). Convergence and extrapolation of spin-state energies for complex \textbf{A: } low-energy spectra for the def2-SVP basis in a (32e, 38o) active space. Plot of the TPSCI variational and PT2 corrected energy as a function of the computed ``PT2" correction. The solid line is the linear fit for variational energy and the dotted line is for ``PT2" corrected energy. Units in wavenumber. $M$=100. (II). Molecular orbitals comprising (32e, 38o) active space. (a) Orbitals of (7e, 10o) Cr local active space. (b) Orbitals of (6e, 6o) bridged hydroxy group local active space. (III). DMRG energy extrapolation for complex \textbf{A} using Fiedler algorithm reordered orbitals for $\chi$=$\infty$. Sweep energy is plotted against the largest discarded weight for each bond dimension in the reverse schedule.
    }
    \label{fig:cr2_pantazis}
\end{figure*}

We have considered two chromium centers with three bridging hydroxyl ligands for complex $\textbf{A}$ to form our active space.
The active space includes 38 molecular orbitals, chosen based on strong overlap with Cr 3d and 4d orbitals, as well as O 2p and 3p orbitals from bridging $\text{OH}^{-}$ ligands, derived from a high-spin ROHF wavefunction optimized for the heptet state.
The active space comprises 13 doubly occupied, 6 singly occupied, and 19 virtual orbitals, resulting in a (32e, 38o) active space and divided into 5 clusters: two Cr clusters with local active spaces of (7e, 10o) and three oxygen clusters with local active spaces of (6e, 6o), as shown in Fig. \ref{fig:cr2_pantazis} (II).

\begin{figure}[ht]
    \centering
    \includegraphics[width=.95\linewidth]{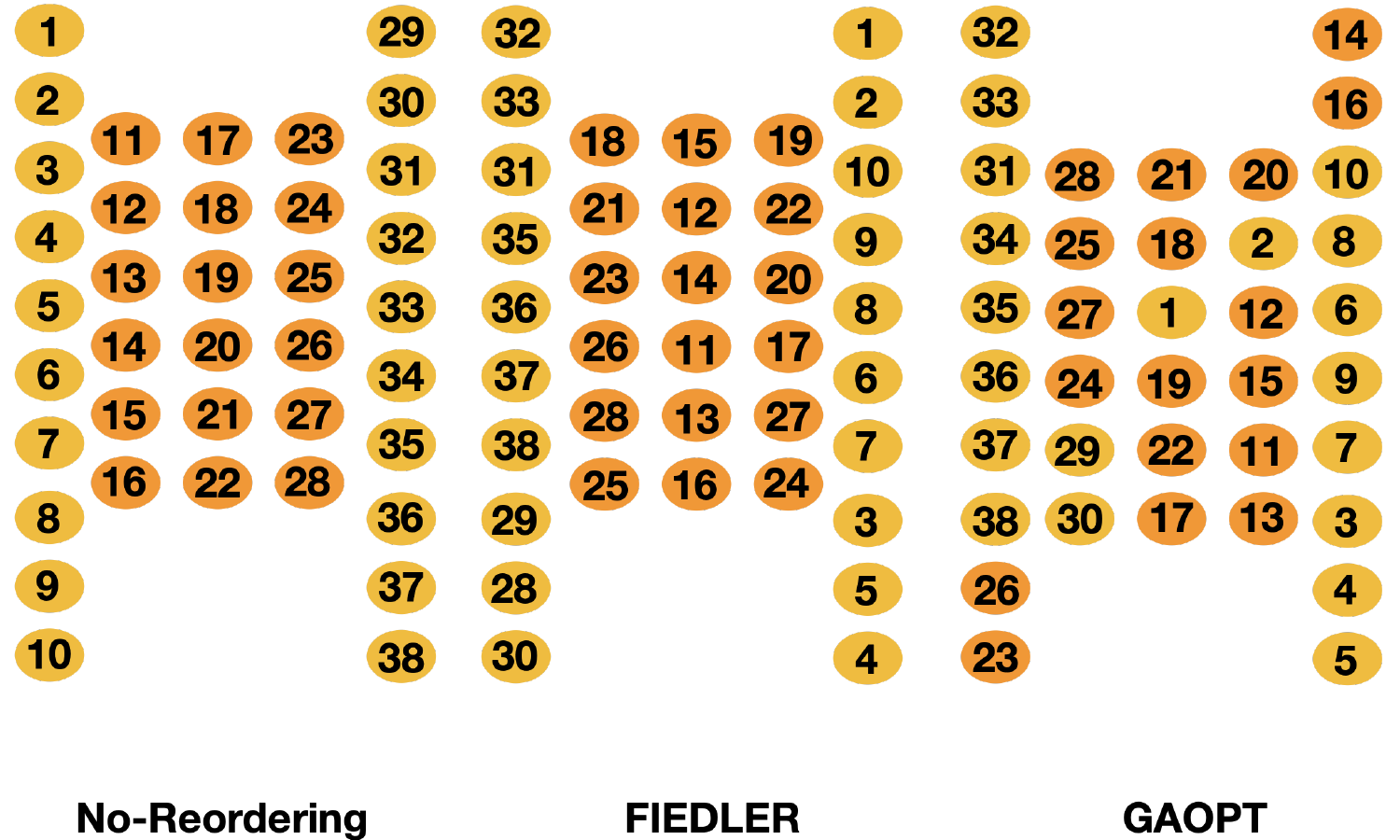}
    \caption{Molecular orbitals comprising (32e, 38o) active space for the complex \textbf{A}. ``No-Reordering" refers to the orbital ordering used in TPSCI calculation. ``FIEDLER" and ``GAOPT" correspond to Fiedler and genetic algorithm sorted orbitals, respectively. }
    \label{fig:cr2_mos_reordering}
\end{figure}

Fig. \ref{fig:cr2_mos_reordering} represents the orbital ordering for the clusters.
``No-Reordering" label is used to show the orbital ordering used in cMF and TPSCI calculation.
Orbitals reordered using Fiedler and GAOPT algorithms are also shown in Fig. \ref{fig:cr2_mos_reordering} labeled as ``FIEDLER" and ``GAOPT", respectively.
Fiedler sorted orbitals do not swap between clusters; however, a significant swap of orbitals occurred in between clusters during ``GAOPT" orbital reordering.

As discussed in Section \ref{theory}, a many-body basis was formed for the sectors of Fock space by diagonalizing the cMF Hamiltonian.
Each Cr cluster spanned Fock sectors from (4e, 10o) to (10e, 10o), while each bridging ligand, $\text{OH}^{-}$ included 7 Fock sectors centered at (6e, 6o).
We form the cluster basis by keeping spin-preserved states for each Fock sector by restricting the number of states with $M$.

\subsection{Complex \textbf{B: }$[\text{Cr}_2\text{O}{(\text{NH}_3})_{10}]^{4+}$}\label{cr2_2_appendix}
\begin{figure*}
    \centering
    \includegraphics[width=0.95\linewidth]{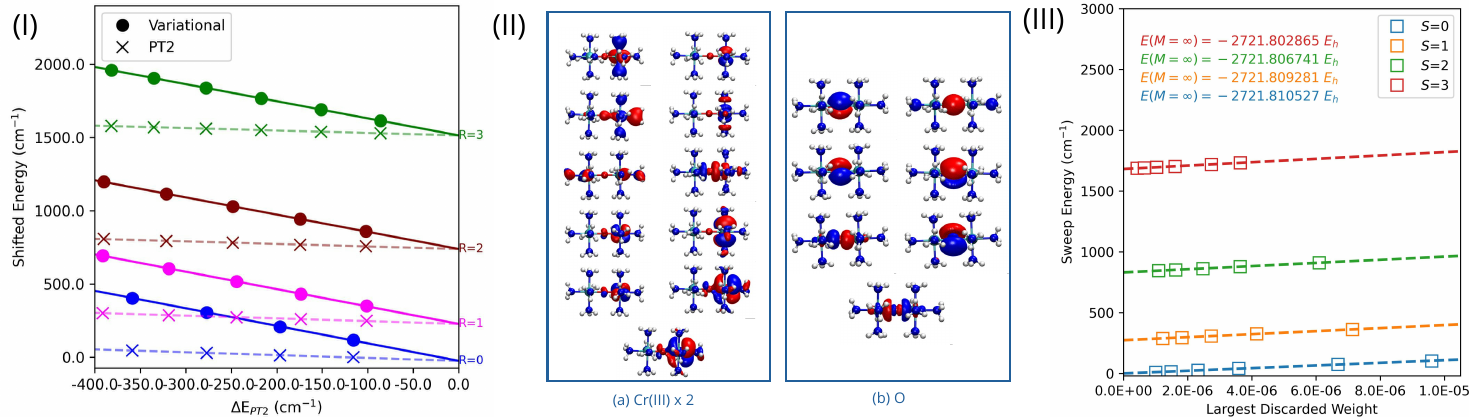}
    \caption{(I). Convergence and extrapolation of spin state energies for complex \textbf{B: } Low-energy TPSCI spectra for the def2-SVP basis in a (26e, 29o) active space. x and y axes, solid and dotted lines represent their usual meanings as in Fig. \ref{fig:cr2_pantazis}. $M$=250. (II). Molecular orbitals comprising (26e, 29o) active space. (a) Orbitals of (9e, 11o) Cr local active space. (b) Orbitals of (8e, 7o) bridged oxygen local active space. (III). DMRG extrapolated energy spectra calculated using Fiedler sorted orbitals. Sweep energy is plotted against the largest discarded weight for each bond dimension ($\chi$) in the reverse schedule to get extrapolated energy at $\chi$=$\infty$. }
    \label{fig:Cr_complex}
\end{figure*}
The active space for complex \textbf{B} is selected similarly to that of complex \textbf{A} as the two chromium centers with only oxygen bridging ligand are considered for active space of complex \textbf{B}.
The active space includes 29 molecular orbitals, chosen based on strong overlap with Cr 3d, 4s, and 4d orbitals, as well as 2p, 3s, and 3p orbitals from bridging oxygen ligand, derived from a high-spin ROHF wavefunction optimized for the heptet state. 
Similar to the active space construction for complex \textbf{A}, MOs are formed to maximally span/overlap selected AOs for complex \textbf{B}.
The resulting active space comprises 10 doubly occupied, 6 singly occupied, and 13 virtual orbitals, resulting in a (26e, 29o) active space, localized to the system and optimized for cMF reference. 
The 29 active orbitals were divided into three clusters: two Cr clusters with local active spaces of (9e, 11o) and one oxygen cluster with local active spaces of (8e, 7o).
The MOs are shown in the Appendix in Fig. \ref{fig:Cr_complex} (II).
For the DMRG calculation, the Fiedler algorithm is used to reorder the orbitals. 
As our active space contains Cr-O-Cr in a highly linear arrangement, this complex is a pseudo-1D system. 
Hence, we would expect that DMRG would be very efficient in tackling this di-chromium complex, serving as a great reference for benchmarking our TPSCI results.


\subsection{Complex \textbf{C: }$[\text{Fe}_2\text{OCl}_6]^{2-}$}\label{fe2_appendix}

\begin{figure*}
    \includegraphics[width=0.95\linewidth]{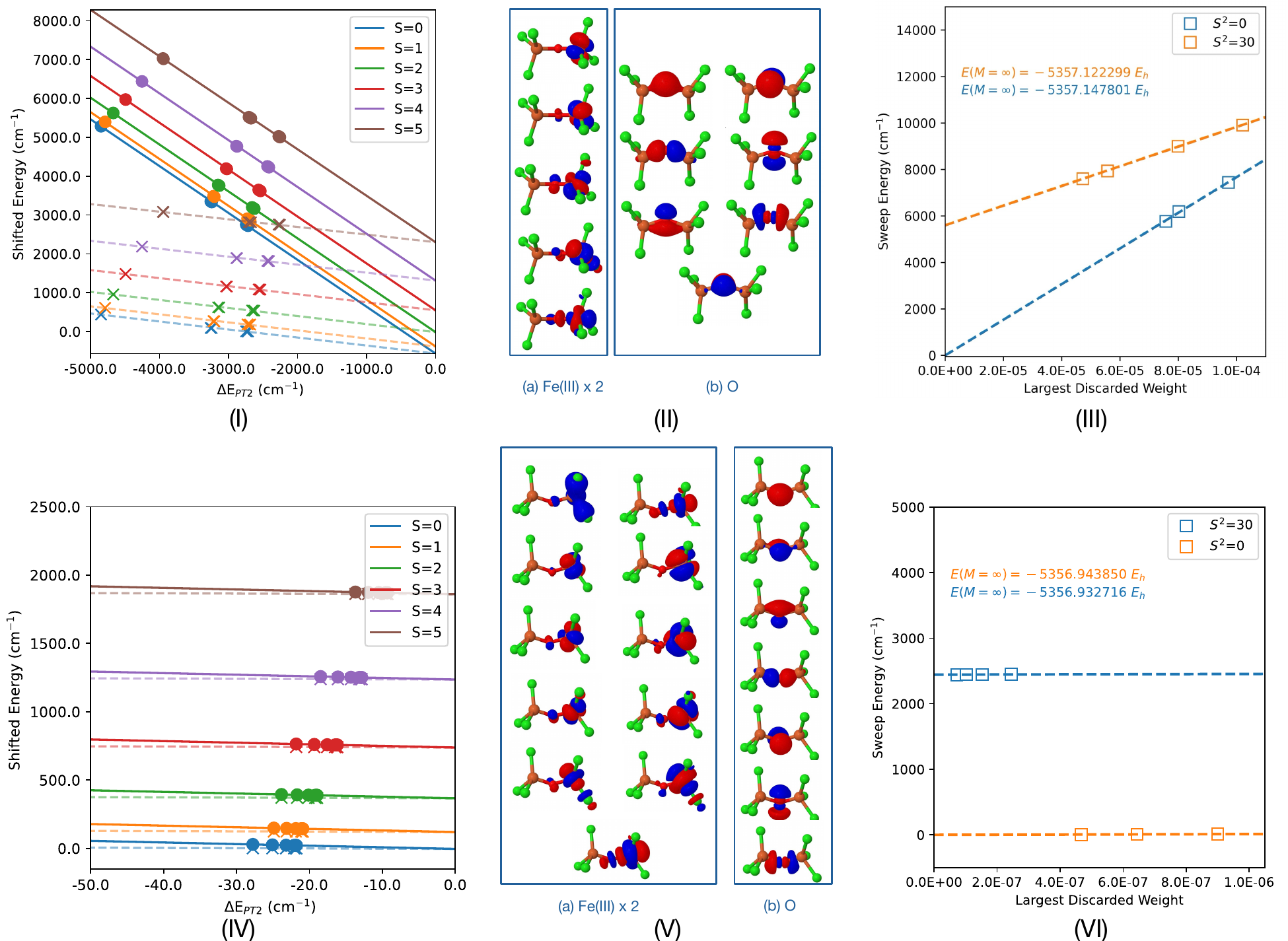}
    \caption{(I). Convergence and extrapolation of spin state energies for complex \textbf{C}: low-energy spectra for the 6-31g* basis in a (42e, 49o) active space.  x and y axes, solid and dotted lines represent their usual meanings as in Fig. \ref{fig:cr2_pantazis}. $M$ = 150. (II). Molecular orbitals comprising the main active space (18e, 17o) of (42e, 49o) active space for complex \textbf{C}.(a) Orbitals of (5e, 5o) Fe local active space. (b) Orbitals of (8e, 7o) bridged oxygen local active space. (III). DMRG extrapolated energy spectra calculated using Fiedler sorted orbitals in (42e, 49o) active space. Sweep energy is plotted against the largest discarded weight for each bond dimension ($\chi$) in the reverse schedule to get extrapolated energy at $\chi$=$\infty$.
    (IV). Low-energy TPSCI spectra for the 6-31g* basis in a (22e, 29o) active space. $M$ = 200.
    (V). Molecular orbitals comprising (22e, 29o) active space for complex \textbf{C}. (a) Orbitals of (7e, 11o) Fe local active space. (b) Orbitals of (8e, 7o) bridged oxygen local active space.
    (VI). DMRG extrapolated energy spectra calculated using Fiedler sorted orbitals in (22e, 29o) active space.}
    \label{fig:fe2ocl6_complex}
\end{figure*}
In a similar way as we chose the (26e, 29o) active space for complex \textbf{B}, Fe 3d, 4s, and 4d orbitals and O 2p, 3s, and 3p orbitals are chosen to be in the active space of complex \textbf{C}. 
The resulting active space, (22e, 29o), contains 6 doubly occupied, 10 singly occupied, and 13 virtual orbitals. 
The 29 active orbitals were divided into three clusters: two Fe clusters with local active spaces of (7e, 11o) and one oxygen cluster with local active spaces of (8e, 7o).

We have taken a different approach to build cluster active spaces for the (42e, 49o) active space.
Here, 3d orbitals are selected for two separate Fe clusters and 2p, and 3p oxygen orbitals are chosen to be in a separate cluster. 
They are projected onto the doubly, singly occupied, and virtual subspaces by keeping the largest singular vectors from each ROHF subspace to choose our initial active space, (18e, 17o). 
Doubly occupied orbitals (Fe 3s, 3p, and O 2s) and virtual orbitals (Fe 4s, 4p, 4d, and O 3d) are projected separately onto different MO subspaces to build separate doubly occupied and virtual MO clusters.
Resulted active space, (42e, 49o) comprises 16 doubly occupied, 10 singly occupied, and 23 virtual orbitals. 
The 49 active orbitals were divided into five clusters: two Fe clusters with local active spaces of (5e, 5o), one oxygen cluster with local active space of (8e, 7o), doubly occupied (24e, 12o), and virtual (0e, 23o) cluster active space.


\subsection{Complex \textbf{D: } \texorpdfstring{$[\text{Fe}_2\text{S}_2{(\text{SCH}_3})_{4}]^{2-}$}{[Fe2S2(SCH3)4]2-} }\label{fe2s2_appendix}
\begin{figure*}
    \centering
    \includegraphics[width=0.95\linewidth]{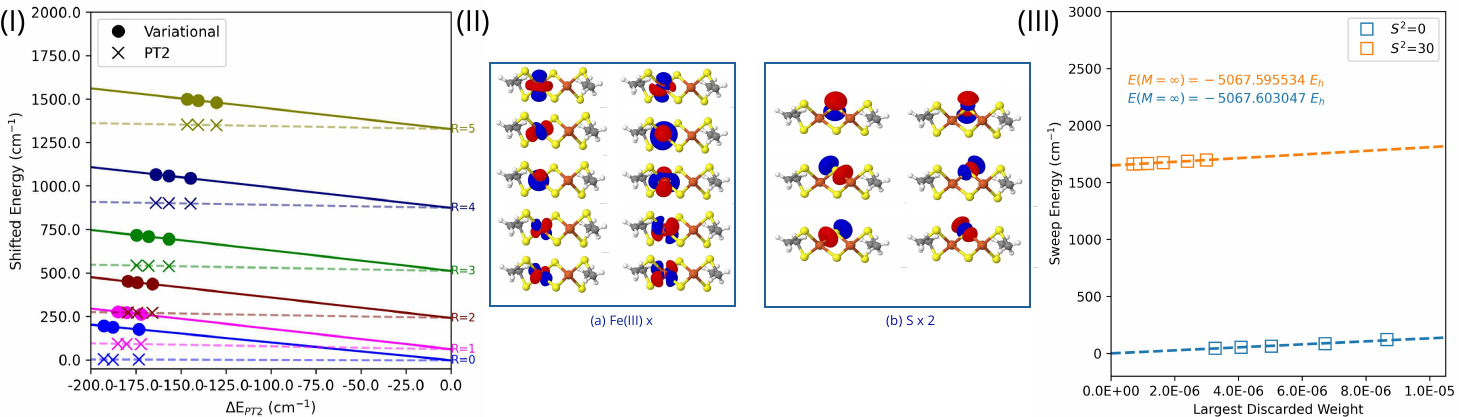}
    \caption{(I). Convergence and extrapolation of spin state energies for complex \textbf{D}: low-energy spectra for the def2-SVP basis in a (22e, 32o) active space.  x and y axes, solid and dotted lines represent their usual meanings as in Fig. \ref{fig:cr2_pantazis}.
(II). Molecular orbitals comprising (22e, 32o) active space for the Iron-sulphur complex \textbf{D}.
    (a) Orbitals of (5e, 10o) Fe local active space. (b) Orbitals of (6e, 6o) bridged sulfur local active space.
    (III). DMRG extrapolated energy spectra calculated using Fiedler sorted orbitals in (22e, 32o) active space. Sweep energy is plotted against the largest discarded weight for each bond dimension ($\chi$) in the reverse schedule to get extrapolated energy at $\chi$=$\infty$.}
    \label{fig:iron-sulphur_complex}
\end{figure*}
We start by considering 3d orbitals localized on each Fe atom as separate clusters and 3p orbitals localized on each bridging S as the other two separate clusters.
We can modify our active space by including MOs that have the most overlap with bridged S orbitals, Fe occupied or virtual AO orbitals.
However, not including the non-bridged S ligands in the active space may miss out on some of the contributions from ligands.
So, it is interesting to see how additions of S ligands in the active space affect the energy levels and related properties.
Including the 3p orbitals of non-bridging S ligands in the active space makes an active space of (46e, 28o).
As we choose our active spaces for the other complexes, the semiautomatic procedure of picking the MOs has been employed.
The resulting active space comprises 18 doubly occupied, 10 singly occupied orbitals.
The 28 active orbitals were divided into eight clusters: two Fe clusters with local active spaces of (5e, 5o) and 6 sulfur clusters with local active spaces of (6e, 3o).

Similarly, to account for double-shell effects, we choose both 3d and 4d orbitals for Fe and 3p and 4p orbitals for the bridging S atoms, resulting in an expanded active space of (22e, 32o), while non-bridging S orbitals are excluded.
The active space comprises 6 doubly occupied, 10 singly occupied, and  16 virtual orbitals.
The 32 active space orbitals were divided into four clusters: two Fe clusters with local active spaces of (5e, 10o) and two sulfur clusters with local active spaces of (6e, 6o).

\subsection{Complex \textbf{E: }$\text{Mn}_2(\mu-\text{O})_2(\text{NH}_3)_8]^{4+}$}\label{mn2_appendix}
 \begin{figure*}
    \centering
    \includegraphics[width=0.95\linewidth]{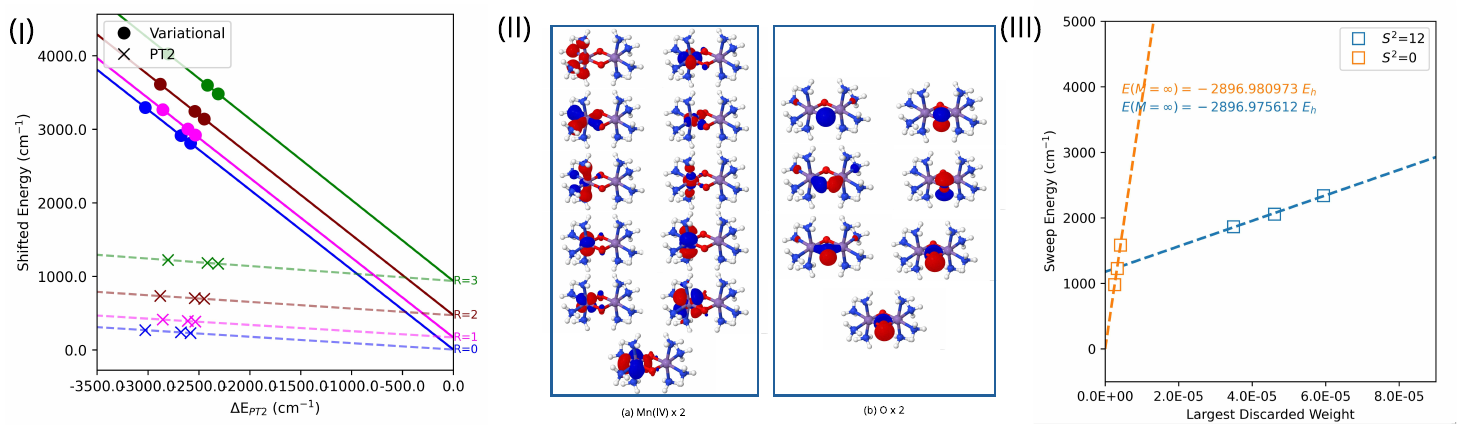}
    \caption{(I). Convergence and extrapolation of spin state energies for complex \textbf{E}: low-energy spectra for the def2-SVP basis in a (34e, 36o) active space for $M$=150. x and y axes, solid and dotted lines represent their usual meanings as in Fig. \ref{fig:cr2_pantazis}.(II). Molecular orbitals comprising (34e, 36o) active space for the Mn dimer complex \textbf{E}. (a) Orbitals of (9e, 10o) Mn local active space. (b) Orbitals of (8e, 7o) bridged oxygen local active space.
    (III). DMRG extrapolated energy spectra calculated using Fiedler sorted orbitals in (34e, 36o) active space. Sweep energy is plotted against the largest discarded weight for each bond dimension ($\chi$) in the reverse schedule to get extrapolated energy at $\chi$=$\infty$.}
    \label{fig:Mn_complex}
\end{figure*}
3d, 4s, and 4d orbitals of Mn and 2p, 3s, and 4p atomic orbitals of oxygen were picked up to form an active space (34e, 36o) (see Fig. \ref{fig:Mn_complex}) (II).
The active space has 14 doubly occupied, 6 singly occupied, and 16 virtual orbitals that get divided into four clusters: two Mn local active spaces of (9e, 10o) and two oxygen local active spaces of (8e, 7o).

\subsection{Complex F: Ni-cubane}
To simulate the Ni-cubane core, we have considered 3d orbitals of Ni centers and 2p and 3p orbitals of bridged oxygen ligands to build an efficient active space.
Each Ni center has two unpaired electrons that lead to eight unpaired electrons which eventually give rise to 19 spin states with $S^2=\{20,12,6,2,0\}$ (see Fig. \ref{fig:Cubane_energy}.
The same approach (see section \ref{activespace}) has been considered to pick the MOs for the active space.
The resulting active space, (56e, 44o) has 24 doubly occupied, 8 singly occupied, and 12 virtual orbitals.
The nickel centers have (8e, 5o) local active spaces, while the oxygen local centers have (6e, 6o) local active spaces.
 \begin{figure*}
    \centering
    \includegraphics[width=0.99\linewidth]{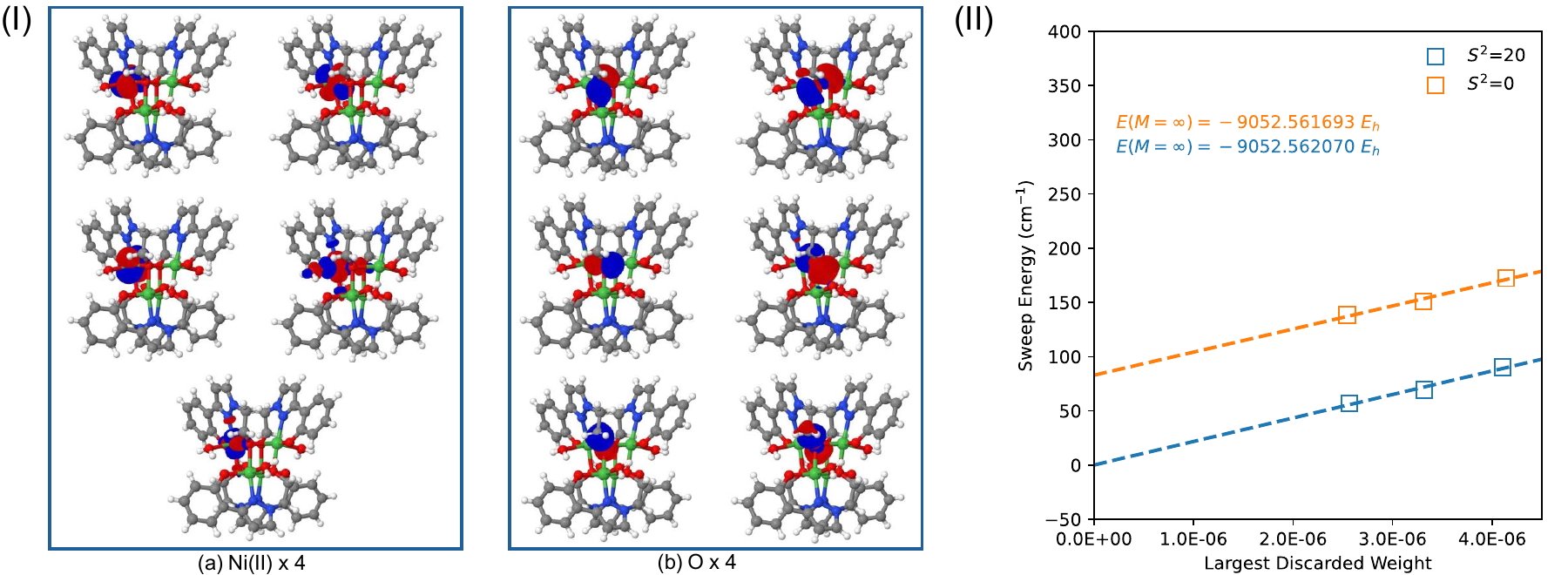}
    \caption{(I). Molecular orbitals comprising (56e, 44o) active space for the nickel cubane complex \textbf{F}. (a) Orbitals of (8e, 5o) Ni local active space. (b) Orbitals of (6e, 6o) bridged oxygen local active space.
    (II).  DMRG extrapolated energy spectra calculated using Fiedler sorted orbitals in (56e, 44o) active space. Sweep energy is plotted against the largest discarded weight for each bond dimension ($\chi$) in the reverse schedule to get extrapolated energy at $\chi$=$\infty$.}
    \label{fig:Cubane_complex}
\end{figure*}
\section{Magnetic exchange coupling constant determination for multi-TM complex.}\label{j_multimetal}

Complex \textbf{F} has four Ni centers with each having a well-defined oxidation state (II) and spin state (S=1).
To get a qualitative description of the low-energy spectrum of this complex, we need a model space.
To ensure that the chosen model space accurately captures the relevant physics, the exact low-energy eigenstates should have significant overlap with the model space.
Consequently, the low-energy spectrum is expected to be primarily composed of 81 spin configurations that provide a well-defined basis for our model.
However, since the Hamiltonian conserves spin, we can further simplify the problem by restricting our analysis to the global $M_s=0$ subspace, reducing the model space to the corresponding 4-dimensional subspace.
Magnetic exchange coupling constant, $J$ between different Ni centers could be calculated from this 4-dimensional subspace following this procedure.
\begin{enumerate}
    \item Perform a single TPSCI calculation for $(S_{max}-1)$ spin state (i.e. 4 spin states).
    \item Diagonalize to get the TPSCI eigenvectors ($V_i$) and eigenvalues ($E_i$): $\hat{H}V_i = E_iV_i$
    \item Project eigenvectors ($V_i$) into the original localized basis (guess vectors or neutral determinants): $P_{\mu \nu} V_{\nu i} = V^\prime_{\mu i}.$
    \item Build overlap matrix ($S_{ij}=V_{\mu i}^{\prime\dagger} V_{\mu j}^\prime$) and use Löwdin’s orthogonalization to reorthogonalize the projected vectors: $\tilde{V} = V^\prime S^{-\frac{1}{2}} $
    \item Build the effective Hamiltonian matrix on the symmetrically orthogonalized neutral determinant basis using the excitation energies obtained from the TPSCI calculation: $H_{eff}=\tilde{V}E_i\tilde{V}^\dagger$
    \item The exchange coupling constant ($J$) between two spin centers, $A$ and $B$ can then be extracted directly from the off-diagonal elements of the effective Hamiltonian:
    \begin{align}\label{j_multicenter}
        J_{\mathrm{AB}}=-\frac{{\tilde{H}}^{\mathrm{AB}}_{eff}}{2 \sqrt{S_{\mathrm{A}} S_{\mathrm{B}}}}
    \end{align}
    
\end{enumerate}

\end{document}